\documentclass[aps,amsmath,amssymb,showpacs,prb,twocolumn,groupedaddress]{revtex4}

\usepackage{graphicx}

\begin{document}


\title{Thermoelectric properties of electrically gated bismuth telluride nanowires}



\author{I.~Bejenari$^1$$^2$}
 \email{bejenari@iieti.asm.md}
  \homepage{http://www.ee.ucr.edu/~ibejenari/}
 \author{V.~Kantser$^2$}
 \email{kantser@iieti.asm.md}
 \author{A.A.~Balandin$^1$}
 \email{balandin@ee.ucr.edu}
\affiliation{
${^1}$ Department of Electrical Engineering and Materials Science and Engineering Program, University of California - Riverside, Riverside, California 92521 U.S.A. \\
${^2}$ Institute of Electronic Engineering and Industrial Technologies, Academy of Sciences of Moldova, Kishinev, MD 2028 Moldova
}

\date{\today}

\begin{abstract}
We theoretically studied  the effect of the perpendicular electric field on the thermoelectric properties of the intrinsic, ${\boldsymbol{n}\--}$type and ${\boldsymbol{p}\--}$type bismuth telluride nanowires with the growth  direction [110]. The electronic structure and the wave functions were calculated by solving self-consistently the system of the Schr\"{o}dinger and Poisson equations using the spectral method. The Poisson equation was solved in terms of the Newton - Raphson method within the predictor-corrector approach. The electron - electron  exchange - correlation interactions were taken into account in our analysis. In the temperature range from 77 to 500 K, the dependences of the Seebeck coefficient, thermal conductivity, electron (hole) concentration, and thermoelectric figure of merit on the nanowire thickness, gate voltage, and excess hole (electron) concentration were investigated in the constant relaxation-time approximation.  The results of our calculations indicate that the external perpendicular electric field can increase the Seebeck coefficient of the bismuth telluride nanowires with thicknesses of 7 - 15 nm by nearly a factor of 2 and enhance ${ZT}$ by an order of magnitude. At room temperature, ${ZT}$ can reach a value as high as 3.4 under the action of the external perpendicular electric field for realistic widths of the nanowires. The obtain results may open up a completely new way for a drastic enhancement of the thermoelectric figure of merit in a wide temperature range.  
\end{abstract}


\maketitle

\section{\label{sec:level1}Introduction}
Nanostructured materials offer a strong potential for improving the thermoelectric figure of merit and efficiency of the thermal-to-electric energy conversion. \cite{Goldsmid, Balandin} Bismuth telluride and its solid solutions (${\text{Bi}_{2-x}\text{Sb}_x \text{Te}_3}$, ${\text{Bi}_2\text{Sb}_{3-y}\text{Se}}$) are among the best known thermoelectric materials for the modern commercial application. These materials possess notable properties such as high anisotropic multi\-valley Fermi surface, small value of the thermal conductivity, and optimal value of the carrier concentration at room temperature (RT). Since the electron dimensional confinement and the  phonon boundary scattering control the electron and phonon transport, bismuth telluride based superlattices and quantum wires have improved thermoelectric properties as compared to bulk bismuth telluride material.  For example, the thermoelectric figure of merit (${ZT}$) of the ${\text{Bi}_{2}\text{Te}_3/\text{Sb}_2 \text{Te}_3}$ superlattices was reported to have the  value of 2.4 while the maximum ${ZT}$ for the bulk bismuth telluride material is less than 1 at room temperature. \cite{Goldsmid, VenkatNature} Due to stronger spatial confinement the bismuth telluride nanowires (NW) are expected to be better than the quantum well superlattices. Bismuth telluride NWs can be fabricated by the electrochemical deposition of the material in the nanopores of anodized alumina membranes, by the Taylor-Ulitovsky technique, and by high pressure injection of the melt into capillaries. \cite{Zhou_APL87, Li_Nanotech17} Using the technique based on translating thin film growth thickness control into planar wire arrays, metallic and semiconductor nanowires with diameters and pitches (center-to-center distances) as small as 8 and 16 nm, respectively, were prepared. \cite{Melosh_Science}

Electric field effect (EFE) is a powerful tool controlling the electrical properties of low dimensional structures. The latest experimental and theoretical studies showed that the EFE, as well as thermal annealing, can significantly improve the thermoelectric properties of Bi, PbSe nanowires and PbTe films. \cite{Butenko_JAP82, Sandomirsky_JAP90, Boukai_am18, Liang_nl9} While dimensional confinement of electrons leads to the modification of their density of states, the Fermi level can be changed due to EFE control of the electron (hole) concentration in NW. In nanowires, EFE on the thermoelectric properties can also be related to the energy spectrum modification and local properties of the electron distribution function. In addition, the applied electric field can cause quantum confinement of carriers. For these reasons, the electric field applied through the side gates camn lead to strong modifications the thermoelectric properties of the nanowires. The analysis of the temperature dependent thermoelectric power provides a reliable way for estimating the majority carrier concentration in nanowires even when the conventional Hall - effect cannot be used. \cite{Lee_APL94}
 
In this paper, we theoretically studied the intrinsic and doped bismuth telluride nanowires grown along [110] direction at RT. The studied NWs have the square cross- sections of ${7\times 7}$ and ${15\times 15\:\text{nm}^2}$. The main objective of this manuscript is to examine a feasibility of a drastic ${ZT}$ enhancement in nanowires via application of the perpendicular electrical field. Specifically, we studied how the change in the electron density of states caused by both the size quantization and electric field influences the thermoelectric properties of NWs. We examined conditions at which the size quantization effects compete with the electric field effect and when both effects cooperate in improving the thermoelectric properties of NWs.  The electron mobility in the NWs is assumed to coincide with the electron mobility in the bismuth telluride bulk material. In our calculations, we used the values of the lattice thermal conductivity, electron (hole) mobility, and carrier effective mass components obtained previously for the bismuth telluride NWs with the growth direction of [110]. \cite{Bejenari_PRB78} The transport parameters were estimated using the semi-classical Boltzman approximation. We used the constant-relaxation-time approximation to calculate the Seebeck coefficient, electrical and thermal conductivities. The application of this approach to the calculation of the transport properties of bismuth telluride NW material has been justified by us elsewhere. \cite{Bejenari_PRB78} In this paper, we also used some of the notations introduced in our previous work. \cite{Bejenari_PRB78}
The rest of the paper is organized as follows. In Section~\ref{sec:level2}, we describe our approach for solving the self-consistent system of the Schr\"{o}dinger and Poisson equations. In Section~\ref{sec:level3}, the dependence of the thermoelectric parameters of bismuth telluride NWs on lateral gate voltage is studied. In Section~\ref{sec:level4}, we consider the local variation of the thermoelectric parameters inside NWs. In Section~\ref{sec:level5}, we discuss the temperature dependence of the thermoelectric parameters. Finally, our conclusions are given in Section~\ref{sec:level6}. In the Appendix, a detailed description of the spectral method, based on the trigonometric functions, and the spectral element method (SEM), based on the Gauss-Lobatto-Lelendre polynomials, and the linearized Poisson equation (LPE) approach are presented. 
\section{\label{sec:level2}Theory}
\subsection{Self-consistent system of the Schr\"{o}dinger and Poisson equations}
We considered a bismuth telluride nanowire connected to the source and drain and separated from the lateral metallic gate by the dielectric ${\text{Si}\text{O}_2}$ layer with the thickness ${d_{\text{Si}\text{O}_2}=300 \:\text{nm}}$. The $y$ axis of the system of coordinates is directed along the nanowire axis. The electronic band structure of the indirect band gap bismuth telluride NW material is given by a parabolic band approximation in the Drable-Wolf six-valley model. \cite{Bejenari_PRB78} In the rhombohedral coordinate system, the main axis of two equivalent electron (hole) ellipsoids is along the direction [110]. For the other 4 electron (hole) equivalent ellipsoids, the main axis is at the angle  ${\pi/3}$ with respect to the direction [110]. The applied electric field is assumed to be parallel to the $z$ axis.  A similar device based on the Bi nanowire with a thickness of 28 nm has been recently studied experimentally. \cite{Boukai_am18} To estimate the nanowire transport properties, we found a solution of the self-consistent system of the Schr\"{o}dinger and Poisson equations taking into account the difference between the effective masses for the non-equivalent ellipsoids 
\begin{widetext}
\begin{equation}
 -\frac{\hbar^2}{2m_x}\frac{\partial^2\Psi}{\partial x^2}-\frac{\hbar^2}{2m_z}\frac{\partial^2\Psi}{\partial z^2}+\left [V_c(x,z)+V_{xc}(x,z)+E_c-e \varphi (x,z)  \right ] \Psi(x,z)=E\Psi(x,z),
\label{eq:1}
\end{equation}
\begin{equation}
- \epsilon \epsilon_0 \Delta \varphi(x,z)=\rho(x,z),
\label{eq:2}
\end{equation}
\end{widetext}
where the local variables $x$ and $z$ range in the intervals ${0\leq x \leq a_x}$  and ${0\leq z \leq a_z}$,  ${E_c}$ is the bottom of the conduction band, ${\varphi}$  is the electrostatic potential, and ${V_c}$ is the confinement potential, which corresponds to an infinitely high  quantum well. The charge density is given by the expression ${\rho = e p_{1D} -e n_{1D}}$, where ${n_{1D}}$ (${p_{1D}}$) is the electron (hole) concentration. ${V_{xc}(x, z)}$ is the electron-electron exchange-correlation energy given by the following expression \cite{Hedin_JPC4, Stern_PRB30}
\begin{widetext}
\begin{equation}
V_{xc}\left(x,z\right)=- \left[ 1+0.03683 r_s \mathbf{ln} \left( 1+ \frac{21}{r_s\left(x,z\right)} \right ) \right ] \left ( \frac{2}{r_s\left(x,z\right) \pi \alpha} \right ) R^{*}_{y},
\label{eq:3}
\end{equation}
\end{widetext}
where  ${\alpha=(4 \pi / 9)^{1/3}}$, ${\alpha^* = 4 \pi \epsilon_0 \epsilon \hbar^2 / e^2m^*}$  is the Bohr radius, ${r_s\left(x,z\right)=(3/4 \pi n\left(x,z\right))^{1/3}/\alpha^*}$, ${R^{*}_{y}=e^2/8 \pi \epsilon_0 \epsilon \alpha^*}$ is the effective Rydberg constant. The dielectric constant of bismuth telluride is ${\epsilon_{BiTe}=90}$ (Ref.\onlinecite{Goldsmid, Sandomirsky_JAP90, Goltsman}). In the exchange-correlation potential, the anisotropic effective mass  ${m^*}$ can be chosen as either a constant average mass ${m^{*}_{xc}=(m_x m_y m_z)^{1/3}}$ or an optically averaged mass ${m^{*}_{xc}=3/(1/m_x +1/m_y +1/m_z)}$ . For both choices of the effective mass, the results of energy calculation are similar. \cite{Shumway_PE8} For the sake of simplicity, in our calculations we use the constant average mass. For the NW growth direction [110], the electron constant average mass is ${m^{e}_{xc}=0.0821 \: m_0}$. The hole average mass is ${m^{h}_{xc}=0.105 \: m_0}$. Due to the small values of the electron (hole) effective masses and a large value of the dielectric constant, the Bohr radius is  ${\alpha^* =57}$ nm and  ${\alpha^* =45}$ nm for electrons and holes, respectively. It is greater by 3 orders of magnitude than the Bohr radius of hydrogen  ${\alpha^* =0.053}$ nm. For the electrons (holes), the Rydberg constant is ${R^{*}_{y}=1.37 \times 10^{-4}}$  eV (${R^{*}_{y}=1.77 \times 10^{-4}}$ eV) which is nearly 2 orders of magnitude less than the Rydberg constant for hydrogen ${R^{*}_{y}=1.1 \times 10^{-2}}$  eV. Since the dielectric constant of bismuth telluride material is high, the Rydberg constant has a rather  small value of about 15 meV compared to the band gap of 480 meV for the NW with a thickness of 7 nm  and the excess hole concentration ${p_{ex}=1 \times 10^{19} \: \text{cm}^{-3}}$. The influence of the electron-electron exchange-correlation processes on the higher-temperature performance of the wires is largely irrelevant, as the values of ${V_{xc}}$ do not exceed 20 meV, which corresponds to about 230 K.  For the intrinsic semiconductor nanowires with a small electron (hole) concentration, we did not take into consideration the local exchange-correlation energy, since the exchange-correlation effects can be neglected in semiconductor heterostructures with the doping concentration of about ${10^{18} \: \text{cm}^{-3}}$ (Ref. \onlinecite{Mohan_JAP95}). After switching the sign in the second derivatives in Eq.~(\ref{eq:1}), we obtain the Schr\"{o}dinger equation for the holes. 

Integrating both sides of Poisson's equation, we obtain the charge neutrality equation for the gated NW in the form \cite{Luscombe_PRB46}
\begin{equation}
n_{1D}(E_F)=p_{1D}(E_F)+a_x \sigma.
\label{eq:4}
\end{equation}
Here, the parameter ${a_x}$ denotes the NW thickness along the x axis. The surface electric charge ${-e \sigma}$ , induced by the gate electric field at the ${\text{SiO}_2/\text{Bi}_2 \text{Te}_3}$ interface, is associated with a Fermi-level-pinning boundary condition. The condition~(\ref{eq:4}) means that the number of electrons and holes that populate the subbands is equal to the number of charge carriers, which occupy the surface states. The surface states do not contribute to the electron transport, because of the strong surface roughness scattering. The applied electric field induces a surface charge on the ${\text{SiO}_2/\text{Bi}_2 \text{Te}_3}$ interface. The induced surface charge doping concentration  ${\sigma}$ can be estimated in the ${\text{gate}-\text{SiO}_2-\text{Bi}_2\text{Te}_3}$ capacitance structure as a function of the gate voltage ${V_g}$ (Ref.~\onlinecite{Boukai_am18})
\begin{equation}
\sigma=\frac{\epsilon_0 \epsilon_{SiO_2}}{ed_{SiO_2}}V_g,
\label{eq:5}
\end{equation}
where  ${\epsilon_{SiO_2}=3.9}$ is the permittivity of the ${\text{SiO}_2}$ layer. The effective doping ${N_d}$ induced by the applied electric field can be estimated from the expression ${N_d=V_g \epsilon_0 \epsilon_{SiO_2} / ed_{SiO_2}a_{NW}}$, where ${a_{NW}}$ is the NW thickness. In the above expression, the gate voltage ${V_g}$ is followed by the factor of ${4.79 \times 10^{16} \: \text{cm}^{-3}/V}$ (${1.02 \times 10^{17} \: \text{cm}^{-3}/V}$) for the NW thickness 15 nm (7 nm). Hence, the NW electron (hole) concentration can be adjusted to the desirable value by tuning the gate voltage. The same effect can be achieved by doping NWs. However, it is accompanied by an undesirable increase of the electron-impurity scattering.  Since the electric field decreases across the dielectric layer, the effective electrostatic potential ${V^{*}_{g}}$ at the ${\text{SiO}_2/\text{Bi}_2 \text{Te}_3}$ interface is given by the expression ${V^{*}_{g}=V_g \epsilon_{SiO_2} a_{NW} / \epsilon_{Bi_2Te_3} d_{SiO_2}}$. The effective gate voltage ${V^{*}_{g}}$ at the ${\text{SiO}_2/\text{Bi}_2 \text{Te}_3}$ interface is less than the gate voltage ${V_g}$ by a factor of 460 (${10^3}$) for the NW thickness of 15 nm (7 nm). The boundary conditions for Poisson's equation are 
\begin{equation}
\varphi(z=0)=0,
\label{eq:6}
\end{equation}
\begin{equation}
\epsilon_0 \epsilon_{Bi_2Te_3} \varphi'(z=a_z)=e \sigma.
\label{eq:7}
\end{equation}
To solve the Schr\"{o}dinger equation, we use the spectral element method (SEM). \cite{Cheng_JCE3, Cheng_JCP227, Cheng_JCE7} For the intrinsic NWs, we solve Poisson's equation after its linearization. \cite{Ridley} In this case, the Schr\"{o}dinger equation can be recasted into the Mathieu equation. \cite{McLachlan, Bejenari_SST19}  For the doped NWs, we solve the nonlinear Poisson equation by means of the Newton-Raphson method using the predictor-corrector approach. \cite{Press, Trellakis_JAP81, Paceli_IEEE} In the next subsection, we describe our method of calculation of the thermoelectric transport coefficients taking into account the local and gate voltage dependences of both the electron wave functions and the energy spectrum.  
\subsection{Thermoelectric transport coefficients}
To calculate thermoelectric parameters, we use the constant relaxation time (${\tau=\tau_0}$) approximation. In this approximation, the electron (hole) mobility ${\mu}$  is assumed to be constant and given as ${\mu = e \tau_0 / m^{*}}$. The electrical conductivity ${\sigma}$ , Seebeck coefficient $S$, and thermal conductivity ${\kappa_{e(h)}}$  for a NW are defined as \cite{Bejenari_PRB78, Ashcroft}
\begin{equation}
\sigma = L^{(0)},
\label{eq:8}
\end{equation}
\begin{equation}
S = - \frac{1}{eT} \frac{L^{(1)}}{L^{(0)}},
\label{eq:9}
\end{equation}
\begin{equation}
\kappa = \kappa_L + \kappa_e + \kappa_h + \kappa_{eh},
\label{eq:10}
\end{equation}
The lattice thermal conductivity measured perpendicular to the trigonal axis, ${\kappa_{L}}$ , is equal to 1.45 W/mK for bulk bismuth telluride  at RT.\cite{Goldsmid_PPS72, Ainsworth_PPSB69} It is inversely proportional to temperature, ${\kappa_{L}=430/T}$, when ${T>50}$ K.\cite{Walker_PPS76} Both the theoretical estimations and measurements of the NW lattice thermal conductivity indicate that it is reduced by an order of magnitude compared to its bulk counterpart.\cite{Khitun_SM26, BorcaTasciuc_APL85} We take into consideration this fact in the calculation of ZT. The gate electric field does not influence the lattice thermal conductivity. The transport matrix elements ${L^{(\alpha)}}$  are defined below in Eqs.~(\ref{eq:12}) and ~(\ref{eq:13}). 
where the electron (hole) part of the thermal conductivity is given by
\begin{equation}
\kappa_{e(h)} = \frac{1}{e^2 T} \left[ L^{(2)} - \frac{ \left( L^{(1)} \right)^2 }{L^{(0)}}  \right ].
\label{eq:11}
\end{equation}
The last term in Eq.~(\ref{eq:10}), ${\kappa_{eh} = T \sigma_e \sigma_h (S_h - S_e)^2 / (\sigma_e + \sigma_h)}$, is attributed to the thermal energy carried by the electron-hole pairs generated at the heated end of the sample and moving to the cold end where they annihilate. The generalized transport matrix element ${L^{(\alpha)}}$ represents the sum over all six valleys of the bismuth telluride material,  ${L^{(\alpha)} = \sum_{i} {L^{(\alpha)}_i}}$. For the bipolar nanowire, the transport matrix element includes both the electron and hole parts, ${L^{(\alpha)} = L^{(\alpha)}_e + L^{(\alpha)}_h}$. In the LPE approach, the form of the transport matrix elements coincides with that given for the bismuth telluride NWs in the absence of the gate voltage \cite{Ashcroft}
\begin{equation}
L^{\alpha}_i = \frac{e^2 \tau}{\pi a_x a_z} \sum_{n,l} \int_{-\infty}^{\infty} \left( - \frac{\text{d}f}{\text{d}k_y}  \right) v^{2}_{i} (k_y) \left[ E_{n,l}(k_y) - E_F \right]^\alpha \text{d} k_y.
\label{eq:12}
\end{equation}
In this case, the electric field effect on the thermoelectric parameters is due to the modification of the energy spectrum caused by the electrostatic potential  ${\varphi(x, z)}$. In the framework of SEM, the local variation of the transport parameters is defined by the electron (hole) wave function.\cite{Luscombe_PRB46} In SEM approximation, the transport matrix elements for each valley have the form
\begin{widetext}
\begin{equation}
L^{\alpha}_i (x,z) = \frac{e^2 \tau}{\pi a_x a_z} \sum_{n,l} |\Psi_{n,l}(x,z)|^2 \int_{-\infty}^{\infty} \left( - \frac{\text{d}f}{\text{d}k_y}  \right) v^{2}_{i} (k_y) \left[ E_{n,l}(k_y) - E_F \right]^\alpha \text{d} k_y.
\label{eq:13}
\end{equation}
\end{widetext}
In this case, the electron concentration for each valley can be written as
\begin{equation}
n_{1D}(x,z) = N^{1D}_{c,v} \sum_{n=1}^{n_{max}} \sum_{l=1}^{l_{max}(n)} \Phi_{-1/2}( \eta_{n, l}) |\Psi_{n,l}(x,z)|^2.
\label{eq:14}
\end{equation}
The thermoelectric parameters are effectively macroscopic parameters. Hence, under an applied lateral electric field, the figure of merit ${ZT}$ and the Seebeck coefficient $S$ are defined by averaging as\cite{Sandomirsky_JAP90}
\begin{equation}
ZT = \frac{\left\langle \sigma(z) S(z) \right\rangle ^2}{\left\langle \sigma(z) \right\rangle \left\langle \kappa(z) \right\rangle} T,
\label{eq:15}
\end{equation}
\begin{equation}
S = \frac{\left\langle \sigma(z) S(z) \right\rangle}{\left\langle \sigma(z) \right\rangle }.
\label{eq:16}
\end{equation}
In the next section, we present the results of the calculation of thermoelectric parameters in dependence on gate voltage in terms of the Spectral Element Method and the Linearized Poisson's Equation approach at room temperature. 
\section{\label{sec:level3}Dependence of the thermoelectric parameters on the gate voltage }
\begin{figure*}
\includegraphics{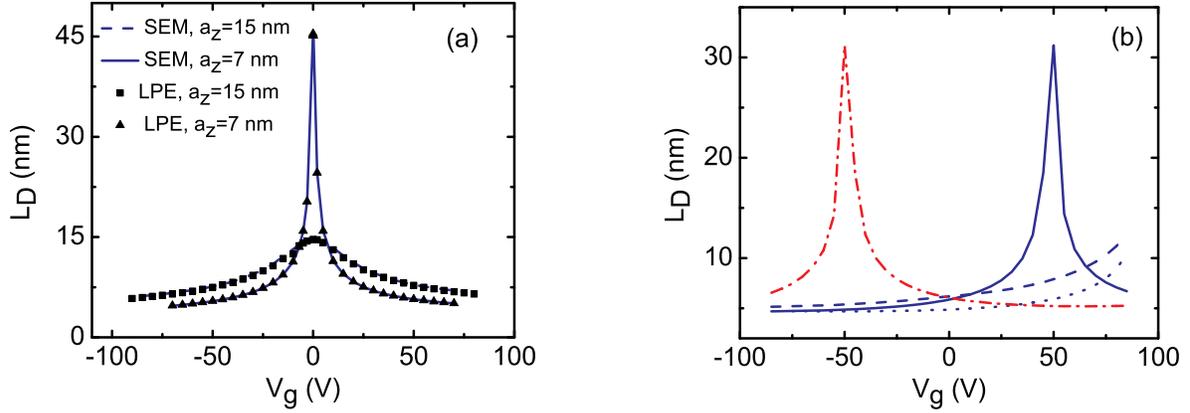}
\caption{\label{fig:1}(Color online) Dependence of the Debye screening length on gate voltage (a) for the intrinsic bismuth telluride NWs with the thicknesses ${a_z=7}$ (solid line) and 15 nm (dashed line), (b) for the doped bismuth telluride NWs with thicknesses of 7 (solid line) and 15 nm (dashed line) at the excess hole concentration ${p_{ex}=5 \times 10^{18} \: \text{cm}^{-3}}$.  The dashed-dotted line (dotted line) corresponds to the NW with a thickness of 7 nm and excess hole concentration ${p_{ex}=1 \times 10^{19} \: \text{cm}^{-3}}$ (excess electron concentration ${n_{ex}=5 \times 10^{18} \: \text{cm}^{-3}}$).  The data obtained by using the LPE approach are depicted by the filled squares and triangles.}
\end{figure*}
For the intrinsic NWs, our calculations show a good agreement between the results obtained in the LPE and SEM approximations. Figure~\ref{fig:1} presents the dependence of the Debye screening length ${L_D}$ for the intrinsic, ${\boldsymbol{n}\--}$type, and  ${\boldsymbol{p}\--}$type bismuth telluride NWs with thicknesses of 7 and 15 nm on the gate voltage ${V_g}$ at room temperature. The Debye screening length decreases with decreasing NW thickness.  The maximum value of the Debye screening length is achieved at ${V_g=0}$ and ${V_g=50}$ V (-50 V) for the intrinsic and ${\boldsymbol{p}\--}$type (${\boldsymbol{n}\--}$type) NWs, correspondingly, with a thickness of 7 nm when the carrier concentration is minimal. There are no peaks in the gate voltage dependence of the ${\boldsymbol{p}\--}$type (${\boldsymbol{n}\--}$type) NWs with a thickness of 15 nm, because their type of conductivity is not changed in the given range of gate voltage. The gate voltage dependence of the Debye screening length for the 15-nm-thick NW is much less than that for the 7-nm-thick NW. For the 15-nm-thick intrinsic NW, ${L_D}$ decreases nearly twice. It is less than the NW thickness in the whole range of the gate voltage. ${L_D}$ decreases by 9 times for the intrinsic NW with a thickness of 7 nm. It is almost equal to the NW thickness at ${V_g = \pm 30}$ V. The ratio of the maximal value of ${L_D}$ to its minimal value is 6 and 1.3, respectively, for the doped NWs (${p_{ex}=5 \times 10^{18} \: \text{cm}^{-3}}$) with thicknesses of 7 and 15 nm.  Therefore, the EFE allows controlling the value of the Debye screening length for both the intrinsic and doped bismuth telluride NWs.
\begin{figure*}
\includegraphics{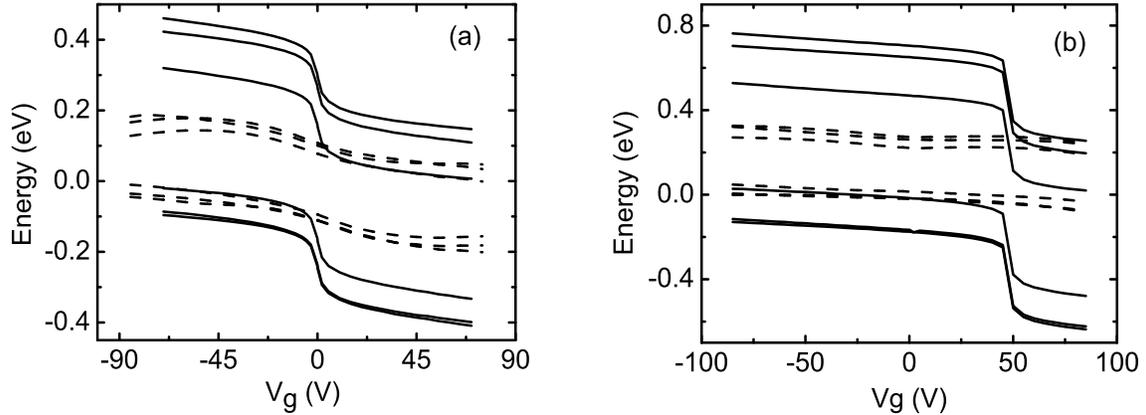}
\caption{\label{fig:2} Dependence of the electron and hole spectrum on gate voltage (a) for the intrinsic and (b) for the ${\boldsymbol{p}\--}$type bismuth telluride NWs with the thicknesses ${a_z=7}$ nm (solid line) and 15 nm (dashed line), when the excess hole concentration is ${p_{ex}=5 \times 10^{18} \: \text{cm}^{-3}}$ at room temperature.}
\end{figure*}
\begin{table*}
\caption{\label{tab:1}  Modification of the electronic structure for the intrinsic bismuth telluride NWs under influence of the gate voltage at room temperature.}
\begin{ruledtabular}
\begin{tabular}{ccccccc}
Thickness, nm & {${V_g}$, V} & ${\Delta E^{e}_{12}}$, meV & ${\Delta E^{e}_{23}}$, meV & ${\Delta E^{h}_{12}}$, meV & ${\Delta E^{h}_{23}}$, meV & ${\Delta E_{g}}$, meV \\ \hline
 7 & 0   & 109  &  33  &  73  &  5  &  322     \\
 7 & 70  & 102 & 38  &   66  &  10  &  340  \\
 15 & 0  & 23  &  7  &   16  &  1   &  171     \\
 15 & 70 & 35  & 10  &   25  &  16  &  160 \\
\end{tabular}
\end{ruledtabular}
\end{table*}

Figure~\ref{fig:2} depicts the dependence of the band diagram of the intrinsic and  ${\boldsymbol{p}\--}$type bismuth telluride NWs with thicknesses of 7 and 15 nm on applied gate voltage with respect to the Fermi energy (${E_F=0}$). The calculations are based on the SEM. Due to the greater size quantization effect, both the separation between the energy subbands and the band gap is greater for the NW with a thickness of 7 nm than those for the NW with a thickness of 15 nm. At the positive gate voltage, the bottom of the conduction band for the NW thickness of 7 nm coincides with that for the NW thickness of 15 nm.  At the negative gate voltage, the top of the valence band for both NW thicknesses is the same. This approaching of the subbands of the NWs with different thicknesses reflects the electric field doping effect, which changes the NW polarity, when the Fermi level approaches the top (bottom) of the valence (conduction) subband.  Since, for the NW thickness of 7 nm, the Debye screening length is greater than the NW thickness at the small values of the gate voltage ${|V_g|<25}$ V (see Fig.~\ref{fig:1}), the dependence of both the electron and hole energy on gate voltage is stronger than that for the NW thickness of 15 nm. For the 7-nm-thick (15-nm-thick) NW, the band gap increases nearly by 5 per cent from 323 meV (172 meV) to 340 meV (183 meV), when the gate voltage ranges in the interval from 0 to 70 V (25 V). For the 15-nm-thick NW, the behavior of the band gap dependence is not monotonic because, at a large value of the gate voltage, the applied electric field influences the size quantization energy levels of electrons and holes in different manner due to the nonlinear electrostatic potential. For the 15-nm-thick NW, the electrostatic potential ${f(V_g)}$ behaves as a parabolic function of gate voltage, ${f(V_g)=a{V_g}^2+bV_g+c}$, where a, b, and c are some constants. It behaves as a linear function of gate voltage, ${f(V_g)=aV_g+b}$, for the 7-nm-thick NW. This different behavior is due to the fact that the Debye screening length for the 7-nm-thick NW is about equal to the NW thickness, while it is about twice less than the thickness of the 15-nm-thick NW (see Fig.~\ref{fig:1}). 
Table~\ref{tab:1} shows that, under the influence of the electric field, the electronic structures for the intrinsic NWs with thicknesses of 7 and 15 nm are modified in different manner. For the intrinsic 7-nm-thick (15-nm-thick) NW, the 1st, the 2nd, and the 3rd electron subbands move up by 149 (74), 156 (62) and 151 (59) meV, correspondingly, when the gate voltage varies from 0 to 70 V. For the 7-nm-thick NW, the spacing between the 1st and the 2nd electron (hole) subbands ${\Delta E^{e}_{12}}$ (${\Delta E^{h}_{12}}$) decreases by 7 meV, while, for the 15-nm-thick NW it increases by 8 meV (9 meV). Also, for the ${\boldsymbol{n}\--}$ and ${\boldsymbol{p}\--}$type NWs, the modification of their electronic structure depends on the NW thickness. 
\begin{figure*}
\includegraphics{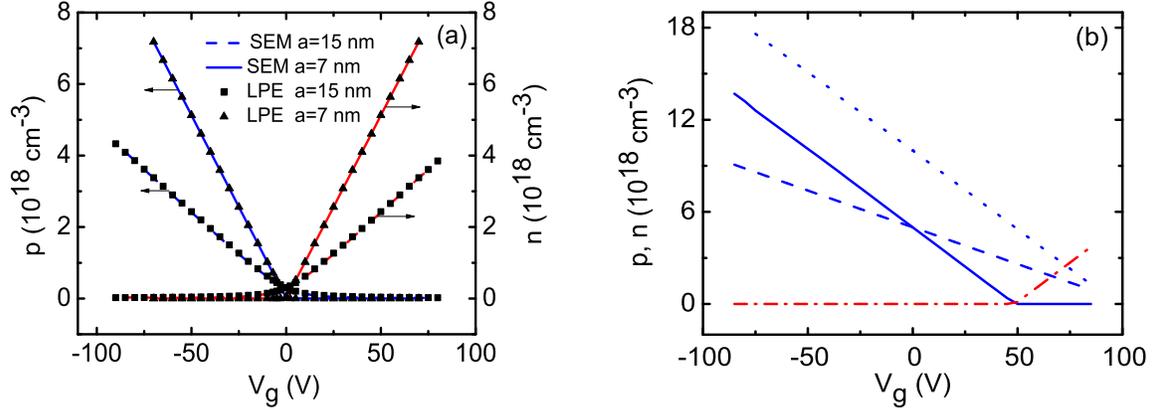}
\caption{\label{fig:3} (Color online) Gate voltage dependence (a) of the electron ($n$) and hole ($p$) concentrations for the intrinsic bismuth telluride NWs with the thicknesses ${a_z=7}$ nm (solid line) and 15 nm (dashed line), (b) of the electron and hole concentrations for the ${\boldsymbol{p}\--}$type NWs with thicknesses of 7 nm (solid line for holes and dashed-dotted line for electrons) and 15 nm (dashed line for holes) when ${p_{ex}=5 \times 10^{18} \: \text{cm}^{-3}}$, at room temperature. The dotted line corresponds to the gate dependence of the hole concentration for the NW thickness of 7 nm when ${p_{ex}=1 \times 10^{19} \: \text{cm}^{-3}}$. The data obtained in the LPE approximation are depicted by the filled squares and triangles.}
\end{figure*}

Figure~\ref{fig:3} shows the linear dependence of the electron and hole concentrations in the intrinsic and ${\boldsymbol{p}\--}$type bismuth telluride NWs on the gate voltage at room temperature. At the zero gate voltage, the charge carrier concentrations in the intrinsic bismuth telluride NWs with the thicknesses of 7 and 15 nm are  ${n=p=3.11 \times 10^{16} \: \text{cm}^{-3}}$  and  ${n=p=3.03 \times 10^{17} \: \text{cm}^{-3}}$, respectively. Under the applied lateral electric field, the carrier concentration increases by two orders of magnitude for the 7-nm-thick intrinsic NW and by one order of magnitude for the 15-nm-thick intrinsic NW. Since the hole effective mass is greater than the electron effective mass, the hole concentration increases faster with the gate voltage compared to the electron concentration. For this reason, for the intrinsic NW with a thickness of 15 nm, the dependence of the electron concentration on the gate voltage is less pronounced compared to that for the holes. At the positive (negative) gate voltage, the electrons (holes) dominate. Therefore, it is possible to control the type of the nanowire conductivity by applying gate voltage. This fact was experimentally confirmed for Bi nanowires.\cite{Boukai_am18} At the excess hole concentration ${p_{ex}=5 \times 10^{18} \: \text{cm}^{-3}}$, the hole concentration can increase nearly twice and three times for the ${\boldsymbol{p}\--}$type NWs with thicknesses of 7 and 15 nm, respectively. The charge carrier concentration diminishes, while the gate voltage dependence of the carrier concentration rises up with a decrease in the NW thickness.  This effect arises from the increased splitting between the energy subbands as a result of a large confinement effect. Thus, the number of subbands involved in the transport decreases with a decrease in the nanowire thickness. For the ${\boldsymbol{n}\--}$type NW, the dependence of the carrier concentration on the gate voltage is reversed compared to that for the ${\boldsymbol{p}\--}$type NW. For the 7-nm-thick NWs with different excess hole concentrations, the dependences of the carrier concentration on the gate voltage are the same.  The effective doping ${N_d}$ induced by the applied electric field is proportional to the gate voltage and inverse proportional to the NW thickness (see discussions below Eq.~\ref{eq:5}). By proper adjusting of the gate voltage, the effective electric doping can compensate the excess holes (electrons).  Hence, the point of neutrality (Vg) for the p(n)-type NW, at which the NW is getting to be intrinsic, depends on both the excess hole (electron) concentration and NW thickness. Therefore, the point of neutrality (${V_g\approx50}$ V) for the 7-nm-thick p-type NW is less than that (${V_g\approx100}$ V) for the 15-nm-thick NW with the same excess hole concentration. The gate voltage dependence of the carrier concentration rises up with a decrease in the NW thickness due to the strong size quantization effect.  Hence, the carrier concentration corresponding to the 15-nm-thick NW is greater (less) than that for the NW with a thickness of 7 nm at the ${V_g=+50}$ V (${V_g=-50}$ V).  

In our calculations, we assumed that the carrier mobility is constant. Since the electrical conductivity is proportional to the carrier concentration, the dependence of the electrical conductivity on the gate voltage Vg is similar with that for the carrier concentration depicted in Fig.~\ref{fig:3}. 
	
Taking into account the gate voltage dependence of the carrier concentration, we calculated the exchange-correlation energy as a function of gate voltage and NW thickness. For the ${\boldsymbol{p}\--}$type NW with a thickness of 7 nm and excess hole concentration ${p_{ex}=5 \times 10^{18} \: \text{cm}^{-3}}$ (${p_{ex}=1 \times 10^{19} \: \text{cm}^{-3}}$), the exchange-correlation energy changes from the value ${V_{xc}=-10}$ meV (-13 meV) at the gate voltage ${V_g=0}$ to ${V_{xc}=-15}$ meV (-17 meV) at the gate voltage ${V_g =-85}$ V. At the excess hole concentration ${p_{ex}=5 \times 10^{18} \: \text{cm}^{-3}}$, the exchange-correlation energy does not change with increasing gate voltage, while, at the excess hole concentration ${p_{ex}=1 \times 10^{19} \: \text{cm}^{-3}}$, it achieves the value ${V_{xc}=-6}$ meV at the gate voltage ${V_g =85}$ V. For the ${\boldsymbol{p}\--}$type NW with a thickness of 15 nm and the excess hole concentration ${p_{ex}=5 \times 10^{18} \: \text{cm}^{-3}}$, the exchange-correlation energy varies from the value ${V_{xc}=-16}$ meV to the values ${V_{xc}=-7}$ meV and -20 meV at the gate voltages ${V_g =85}$ V and ${V_g =-85}$ V, correspondingly. For the ${\boldsymbol{n}\--}$type NW with a thickness of 7 nm and the excess electron concentration ${p_{ex}=5 \times 10^{18} \: \text{cm}^{-3}}$, the exchange-correlation energy varies from the value ${V_{xc}=-10}$ meV to the value ${V_{xc}=-15}$ meV at the gate voltage ${V_g =85}$ V, while ${V_{xc}}$ does not change with decreasing gate voltage.   
\begin{figure*}
\includegraphics{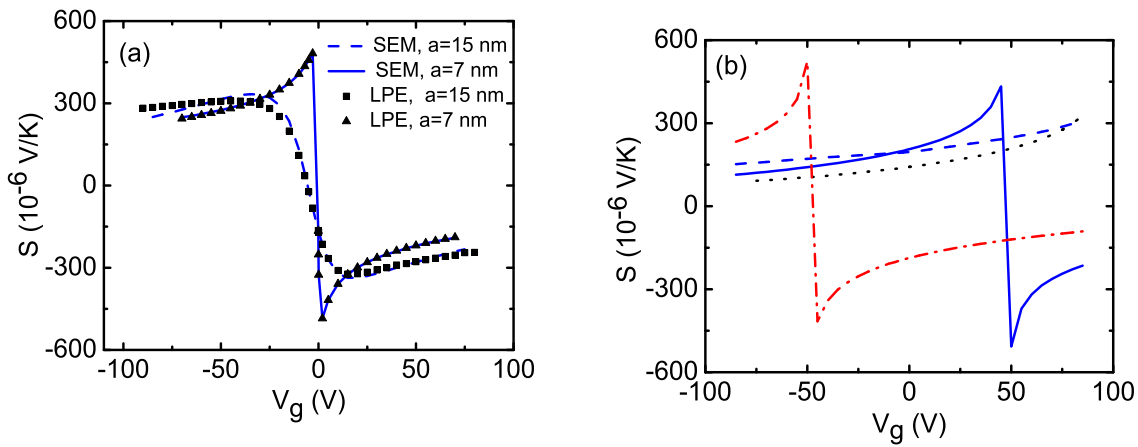}
\caption{\label{fig:4} (Color online) Gate voltage dependence of the Seebeck coefficient ($S$) (a) for the intrinsic and (b) for the ${\boldsymbol{p}\--}$type bismuth telluride NWs with the thicknesses ${a_z=7}$ nm (solid line) and ${a_z=15}$ nm (dashed line) when ${p_{ex}=5 \times 10^{18} \: \text{cm}^{-3}}$ at room temperature. The dashed-dotted line (dotted line) corresponds to the $S$ for the ${\boldsymbol{n}\--}$type (${\boldsymbol{p}\--}$type) bismuth telluride NW with a thickness of 7 nm when ${n_{ex}=5 \times 10^{18} \: \text{cm}^{-3}}$ (${p_{ex}=1 \times 10^{19} \: \text{cm}^{-3}}$).  The data obtained in the LPE approach are depicted by the filled squares and triangles.}
\end{figure*}

Figure~\ref{fig:4} shows the dependence of the nanowire Seebeck coefficient ($S$) on gate voltage at room temperature. At the zero gate voltage, the Seebeck coefficient is -167 and -252  V/K for the intrinsic NWs with thicknesses of 15 and 7 nm, correspondingly. For the intrinsic NW with a  thickness of 7 nm, the maximum absolute value of the Seebeck coefficient approaches the value of 478  V/K (494  V/K) at the gate voltage of -3 V (2 V). For the 15-nm-thick intrinsic NW, the maximum absolute value of 333  V/K (344  V/K) is achieved at the greater gate voltage of -29 V (15 V). Since the electron mobility, ${1200\: \text{cm}^2/\text{Vs}}$, in the binary-bisectrix plane of the bismuth telluride material is twice as large as the hole mobility, ${510\: \text{cm}^2/\text{Vs}}$ (Ref.~\onlinecite{Bejenari_PRB78}), the maximum absolute values of the Seebeck coefficient corresponding to the opposite gate polarities are achieved at different values of ${V_g}$. For the intrinsic NWs, the EFE causes the Seebeck coefficient to increase as much as twice. The confinement effect on the electron density of states leads to an increase in the NW Seebeck coefficient, while the increase in the carrier concentration causes a decrease in $S$ (Ref.~\onlinecite{Bejenari_PRB78}). The larger electron density of states ${D(E)}$ for NWs means a small Fermi-Dirac distribution function ${f(E_c-E_F)}$ and thus a larger difference ${(E_c-E_F)}$. As a result, the Seebeck coefficient for nanowires is higher than that of their bulk counterparts, whereas the electrical conductivity is the same.  Therefore, the absolute value of the Seebeck coefficient strongly increases at the low gate voltage due to the increase in the confinement effect. If the gate voltage continues to increase, ${|S|}$ decreases because the carrier concentration increases with the gate voltage. In the gate voltage interval from -90 to 90 V, the ${\boldsymbol{p}\--}$type NW with a thickness of 15 nm (7 nm) and the excess hole concentration ${p_{ex}=5 \times 10^{18} \: \text{cm}^{-3}}$ (${p_{ex}=1 \times 10^{19} \: \text{cm}^{-3}}$) does not change its type of conductivity (see Fig.~\ref{fig:3}(b)); hence, the dependence of the Seebeck coefficient of such NW on the gate voltage is monotonic. It is increased by 60 per cent (by 1.3 times) with gate voltage. For the 7-nm-thick ${\boldsymbol{p}\--}$type (${\boldsymbol{n}\--}$type) NW with the excess hole (electron) concentration ${5 \times 10^{18} \: \text{cm}^{-3}}$, the absolute value of the Seebeck coefficient ${|S|}$ varies from the value of 187  V/K (206  V/K) at the zero gate voltage to the value of 523  V/K  (507  V/K) at ${V_g=-50}$ V  (50 V) when the NW type of conductivity changes (see Fig.~\ref{fig:3}(b)). In this case, the EFE increases ${|S|}$ by 2.8 (2.5) times. Thus, both the confinement effect and the electric field effect considerably improve the Seebeck coefficient of NW.
\begin{figure*}
\includegraphics{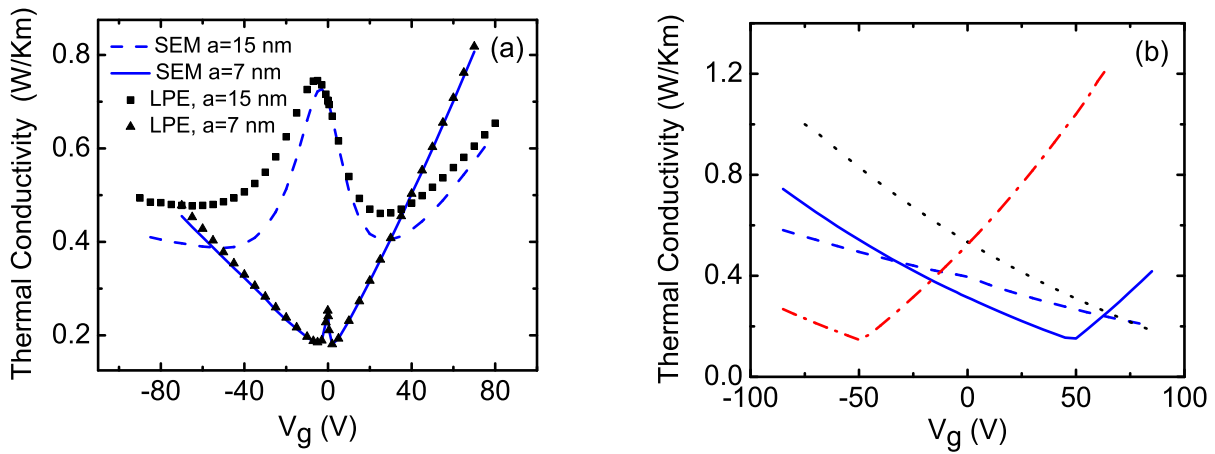}
\caption{\label{fig:5} (Color online) Dependence of the thermal conductivity on gate voltage (a) for the intrinsic and (b) for the ${\boldsymbol{p}\--}$type bismuth telluride NWs with thicknesses of 7 nm (solid line) and 15 nm (dashed line) when ${p_{ex}=5 \times 10^{18} \: \text{cm}^{-3}}$ at room temperature. The dashed-dotted line (dotted line) corresponds to the thermal conductivity of the ${\boldsymbol{n}\--}$type (${\boldsymbol{p}\--}$type) bismuth telluride NW with a thickness of 7 nm when ${n_{ex}=5 \times 10^{18} \: \text{cm}^{-3}}$  (${p_{ex}=1 \times 10^{19} \: \text{cm}^{-3}}$).  The data obtained in the LPE approximation are depicted by the filled squares and triangles.}
\end{figure*}

Figure~\ref{fig:5} depicts the gate voltage dependence of the thermoelectric conductivity for the NWs with thicknesses of 7 and 15 nm. For the intrinsic bismuth telluride NW with a thickness of 15 nm, a small disagreement between the LPE and SEM data appears owing to the great number of the subbands involved in the transport and the rather complex mathematical expression corresponding to the electron part of the thermal conductivity. The dependence of the intrinsic NW thermal conductivity on the gate voltage is not monotonic. The confinement effect on the electron density of states leads to an increase in both the NW Seebeck coefficient and the thermal conductivity, while the increase in the carrier concentration causes their decrease.\cite{Bejenari_PRB78}   At the zero gate voltage, the thermal conductivity is 0.248 W/Km (0.702 W/Km) for the NW with a thickness of 7 nm (15 nm). The minimum value of the NW thermal conductivity of 0.18 W/Km (0.39 W/Km) is achieved at the gate voltage of 2 V (-50 V). For the intrinsic NW with thicknesses of 7 nm and 15 nm, the EFE decreases the thermal conductivity by 30 and 45 per cent, respectively. The interval between the extreme points of the gate voltage for the 7-nm-thick NW is less than that for the 15-nm-thick NW because of the strong dependence of the carrier concentration on gate voltage for the NW with small thickness (see Fig.~\ref{fig:3}).   For the 7-nm-thick NW of the ${\boldsymbol{p}\--}$type (${\boldsymbol{n}\--}$type) at the excess hole (electron) concentration of ${p_{ex}=5 \times 10^{18} \: \text{cm}^{-3}}$ (${n_{ex}=5 \times 10^{18} \: \text{cm}^{-3}}$), the minimal thermal conductivity of 0.152 W/Km (0.147 W/Km) is achieved at the gate voltage of -50 V (50 V) when the type of the NW electrical conductivity changes into the opposite one. Therefore, the EFE halves the thermal conductivity of the 7-nm-thick (15-nm-thick) ${\boldsymbol{p}\--}$type NW when applying gate voltage. The difference between the extreme values of the ${\boldsymbol{p}\--}$ and ${\boldsymbol{n}\--}$type NW thermal conductivities is due to the difference between the electron and hole effective masses as well as the difference in their mobility. For the 7-nm-thick (15-nm-thick) ${\boldsymbol{p}\--}$type NW at ${p_{ex}=5 \times 10^{18} \: \text{cm}^{-3}}$, the gate voltage dependence of the thermal conductivity is monotonic, because the type of NW electrical conductivity does not change in the given gate voltage interval.
\begin{figure*}
\includegraphics{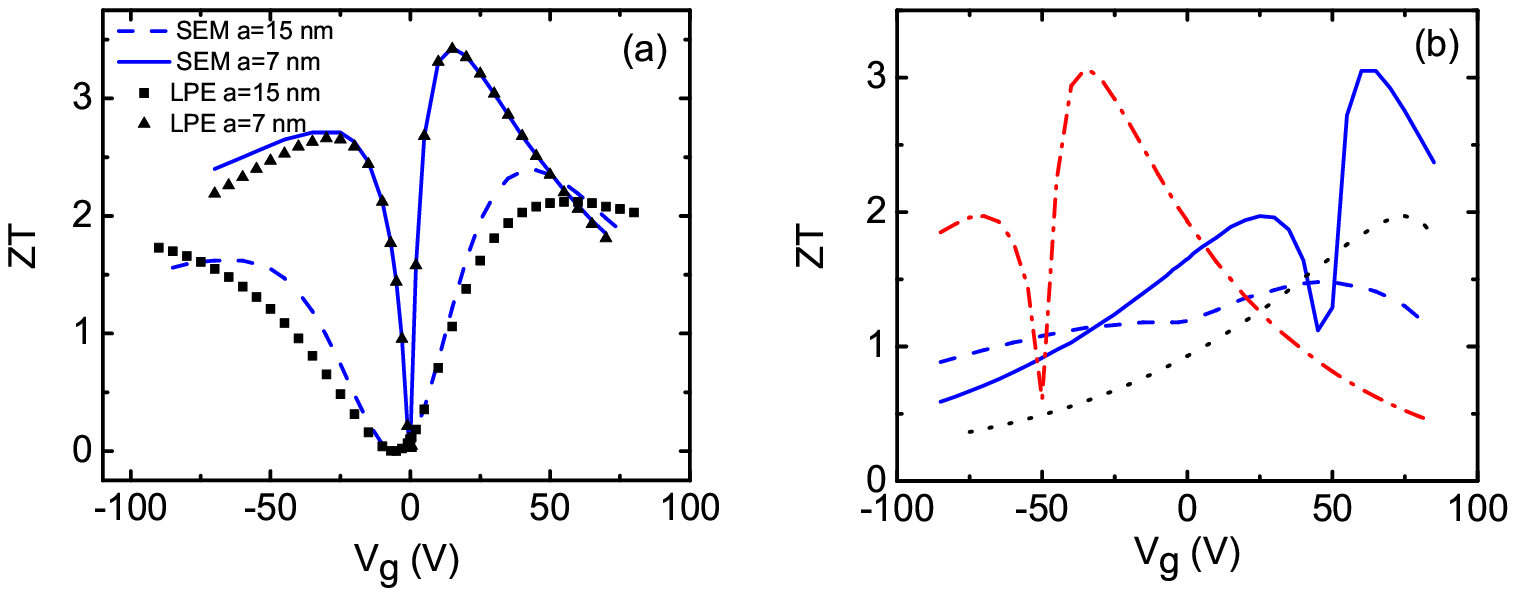}
\caption{\label{fig:6} (Color online) The gate voltage dependence of the figure of merit (${ZT}$) (a) of the intrinsic and (b) of the ${\boldsymbol{p}\--}$type bismuth telluride NWs with the thicknesses ${a_z=7}$ nm (solid line) and ${a_z=15}$ nm (dashed line) when ${p_{ex}=5 \times 10^{18} \: \text{cm}^{-3}}$ at room temperature. The dashed-dotted line (dotted line) corresponds to the ${ZT}$ of the ${\boldsymbol{n}\--}$type (${\boldsymbol{p}\--}$type) bismuth telluride NW with a thickness of 7 nm when ${n_{ex}=5 \times 10^{18} \: \text{cm}^{-3}}$  (${p_{ex}=1 \times 10^{19} \: \text{cm}^{-3}}$).  The data obtained in the LPE approximation are depicted by the filled squares and triangles.}
\end{figure*}

Figure~\ref{fig:6} presents the gate dependence of the thermoelectric figure of merit (${ZT}$) at room temperature. At the zero gate voltage, for the intrinsic NWs with a thickness of 7 nm (15 nm), the figure of merit is 0.065 (0.099). At the gate voltage of 15 V (42 V), the maximum value of the figure of merit is as high as 3.4 (2.4). For the opposite gate polarity, when the holes dominate, the extreme value of the figure of merit is lower than that corresponding to the positive gate voltage because of the difference between the hole mobility and the electron mobility, as well as due to the difference in the effective masses of electrons and holes. For the 7-nm-thick ${\boldsymbol{n}\--}$ and ${\boldsymbol{p}\--}$type NWs, the maximum value ${ZT=3.1}$ is less than that value for the intrinsic NW, ${ZT=3.4}$. For the 7-nm-thick NW, both the figure of merit and the Seebeck coefficient have two extreme values due to the change of the NW conductivity type. When the electrons dominate in the 7-nm-thick ${\boldsymbol{p}\--}$type NW, the figure of merit of 3.1 at ${V_g=60}$ V is greater than its value of 1.97 at ${V_g=25}$ V by 60 per cent (see Fig.~\ref{fig:6}(b)). At the room temperature, for the 7-nm-thick ${\boldsymbol{p}\--}$type NW, the maximum value of the figure of merit 3.1 is greater than the corresponding value ${ZT=1.65}$ at the zero gate voltage by 90 per cent. Since, for the 15-nm-thick NW at ${p_{ex}=5 \times 10^{18} \: \text{cm}^{-3}}$ and the 7-nm-thick NW at ${p_{ex}=1 \times 10^{19} \: \text{cm}^{-3}}$, the type of the electrical conductivity does not change within the whole interval of the gate voltage, their figure of merit gradually increases from 1.19 and 0.93 at the zero gate voltage to 1.48 and 1.97 at ${V_g=45}$ V and ${V_g=75}$ V, respectively. Therefore, the applied lateral electric field improves the NW thermoelectric properties considerably.
\section{\label{sec:level4} LOCAL VARIATION OF THE THERMOELECTRIC PARAMETERS }
In this section, the  local variation of thermoelectric parameters corresponding to the maximum value of the figure of merit at room temperature is discussed. The calculations are based on the SEM.  For the gated bismuth telluride NWs, the local (maximal and minimal) values of thermoelectric parameters as well as their values averaged over the NW cross-sections are listed in Table~\ref{tab:2}. The electron (hole) concentration tends to zero at the NW boundary, because we impose the zero boundary condition on the electron (hole) wave function here. For the 7-nm-thick (15-nm-thick) intrinsic NW, the averaged value of the electron concentration is ${1.5 \times 10^{18} \: \text{cm}^{-3}}$  (${1.9 \times 10^{18} \: \text{cm}^{-3}}$) at the gate voltage  ${V_g=15}$ V (${V_g=42}$ V). The electron distribution in the NW cross section is more gradual for the small NW thickness because of the strong size quantization effect and the small number of subbands participating in the transport for such NWs. As a result of the strong size quantization, for the less NW thickness, the greater gate voltage should be applied to modify the electron dimensional probability density function. For the 7-nm-thick (15-nm-thick) intrinsic NW, the Debye screening length 9.5 nm (8.2 nm) is less (greater) than the NW thickness at a gate voltage ${V_g=15}$ V (${V_g=42}$ V) (see Fig.~\ref{fig:1}(a)). It also causes the uniform (non-uniform) charge carrier distribution in the 7-nm thick (15-nm-thick) NW cross section. For the ${\boldsymbol{n}\--}$ and ${\boldsymbol{p}\--}$type NWs with a thickness of 7 nm, their optimal profiles of the electron distribution in the NW cross section are similar at the gate voltages ${V_g=-35}$ V and ${V_g=60}$ V, when the figure of merit is maximal (see Fig.~\ref{fig:6}(b)). The corresponding Debye screening length, 10 nm (11 nm), is greater than the thickness of the ${\boldsymbol{n}\--}$type (${\boldsymbol{p}\--}$type) NW (see Fig.~\ref{fig:1}(b)). For the ${\boldsymbol{n}\--}$type NW, the optimal profile of the charge carrier distribution is obtained at less gate voltage compared to that for the ${\boldsymbol{p}\--}$type NW.  Hence, the gate voltage enables charge carrier concentration for the NWs to be manipulated. The local variation of the exchange-correlation energy is similar to that of the charge carrier concentration.
\begin{table*}
\caption{\label{tab:2}  Thermoelectric parameters of the gated bismuth telluride NWs at room temperature.}
\begin{ruledtabular}
\begin{tabular}{ccccccc}
 &\multicolumn{2}{c}{Intrinsic}&\multicolumn{3}{c}{${\boldsymbol{p}\--}$type} & ${\boldsymbol{n}\--}$type\\ \hline
Thickness, nm  &  7  &  15 &   7 &   15 &   7 &  7      \\
 ${V_g}$, V   &  15 &  42  &  60 &   45 &  75 &  -35     \\
 ${p_{ex}}$, ${n_{ex}}$, ${10^{18} \: \text{cm}^{-3}}$ 
               &  -  &  -  & 5.0 &  5.0 &  10 &  5.0      \\
 ${L_D}$, nm &   9.5 & 8.5 &  11 &  7.6 &   8 &  10    \\
 ${n_{max}}$, ${10^{18} \: \text{cm}^{-3}}$ & 
                 2.2 & 5.6 & 1.9 &   -  &  -  &  2.1   \\
 ${<n>}$, ${10^{18} \: \text{cm}^{-3}}$ &  
    		         1.5 & 1.9 & 1.2 &  -   &   - &  1.4  \\
 ${n_{min}}$, ${10^{18} \: \text{cm}^{-3}}$ &  
                0.53 & 0   & 0.21 & -   &   - &  0.39  \\
${p_{max}}$, ${10^{18} \: \text{cm}^{-3}}$ & 
${8.2\times10^{-3}}$ & 0.64 & -   & 7.3 & 4.4 &  -    \\
 ${<p>}$, ${10^{18} \: \text{cm}^{-3}}$ &  
 ${5.6\times10^{-4}}$ & 0.025 & - & 2.8 & 2.3 &  -   \\
 ${p_{min}}$, ${10^{18} \: \text{cm}^{-3}}$ &  
                   0 &  0  &  -   & 0.31 & 0.25 &  -  \\  
 ${S_{max}}$, ${10^{-6}}$ V/K & 
                 340 & 381 & 322  &  374 &  296 &  304 \\
 ${<S>}$, ${10^{-6}}$ V/K &  
                 324 & 306 & 319  &  242 &  276 &  301 \\
 ${S_{min}}$, ${10^{-6}}$ V/K &  
                 320 & 257 & 318  &  213 &  274 &  300 \\
 ${\kappa_{max}}$, W/mK &  
               0.313 & 0.704&0.272 &0.453& 0.280& 0.286 \\
 ${<\kappa>}$, W/mK &  
               0.273 & 0.426&0.223 &0.278& 0.217& 0.240 \\
 ${\kappa_{min}}$, W/mK &  
               0.204 & 0.323&0.160 &0.172& 0.157& 0.173 \\
 ${<ZT>}$ &    3.4   &  2.4 &  3.1 & 1.5 &  2.0 & 3.1   \\
\end{tabular}
\end{ruledtabular}
\end{table*}

For the intrinsic NW with a thickness of 7 nm (15 nm) at ${V_g=15}$ V (${V_g=42}$ V), the averaged value of the Seebeck coefficient is -324  V/K (-306  V/K).  For the intrinsic 15-nm-thick NW, the minimal value of ${|S|}$, 257  V/K, is achieved at the local point corresponding to the extreme value of the NW electron concentration (see Fig.~\ref{fig:7}(a)). When the electron and hole local concentrations are the same, the corresponding absolute value of the local Seebeck coefficient, 381  V/K, is maximal. For the 7-nm-thick intrinsic and doped NWs with the excess electron (hole) concentration ${5 \times 10^{18} \: \text{cm}^{-3}}$, the local variation of the Seebeck coefficient is weak, which is coherent with the local variation of the NW carrier concentration. The Seebeck coefficient is negative for both the ${\boldsymbol{p}\--}$type and the ${\boldsymbol{n}\--}$type bismuth telluride NWs with a thickness of 7 nm when ${p_{ex}=5 \times 10^{18} \: \text{cm}^{-3}}$ and ${n_{ex}=5 \times 10^{18} \: \text{cm}^{-3}}$, respectively, at the applied gate voltages 60 and -35 V.  Whereas, $S$ is positive for the ${\boldsymbol{p}\--}$type NWs with thicknesses of 7 and 15 nm when ${p_{ex}=1 \times 10^{19} \: \text{cm}^{-3}}$ and ${p_{ex}=5 \times 10^{18} \: \text{cm}^{-3}}$, correspondingly, at ${V_g=75}$ and 45 V. For the doped NWs, $S$ achieves its maximal value when the carrier concentration tends to its minimal value. Hence, the applied lateral electric field allows controlling the profile of the Seebeck coefficient in the cross section of the 15-m-thick NW.

For the 7-nm-thick NW, the local variation of the thermal conductivity is more gradual compared to that for the 15-nm-thick NW because of the same reasons described above that justify the gradual local variation of the charge carrier distribution.  For the 7-nm-thick ${\boldsymbol{n}\--}$ and ${\boldsymbol{p}\--}$type NWs, at the excess electron and hole concentrations ${n_{ex}=5 \times 10^{18} \: \text{cm}^{-3}}$ and ${p_{ex}=1 \times 10^{19} \: \text{cm}^{-3}}$, respectively, at the gate voltages ${V_g=-35}$ V and ${V_g=75}$ V, the local variations of their thermal conductivities are the same.  As a result, the averaged values of their thermal conductivities, 0.24 and 0.22 W/mK, are very close.  The increase in the local thermal conductivity is related to the increase in the charge carrier concentration at the given point of the NW cross section. The thermal conductivity is less for the NW thickness of 7 nm than that for the NW with a thickness of 15 nm due to the enhancement of the dimensional confinement with a decrease in the NW thickness. Therefore, the electric field effect permits controlling profile of the thermal conductivity in the NW cross section.
\section{\label{sec:level5} TEMPERATURE DEPENDENCE OF THE THERMOELECTRIC PARAMETERS }
\begin{figure*}
\includegraphics{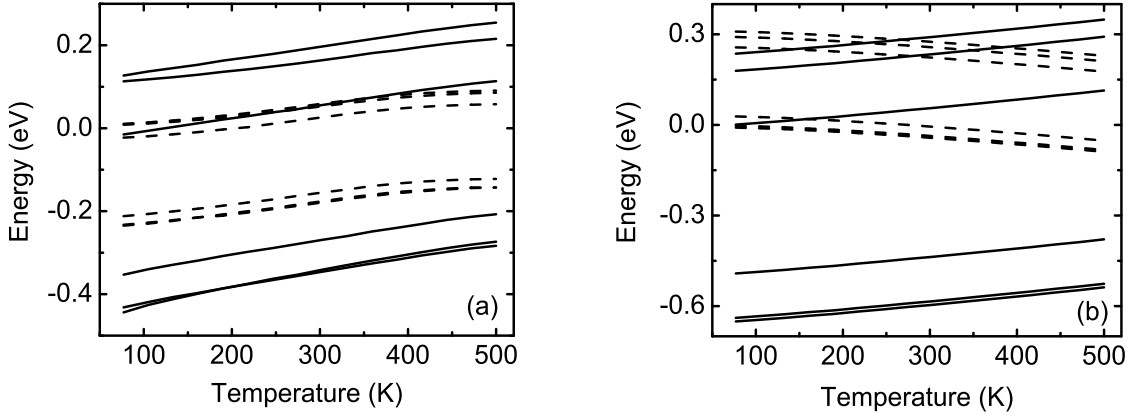}
\caption{\label{fig:7} Temperature dependence of the first three electron and hole subbands (a) for the intrinsic and (b) for the  ${\boldsymbol{p}\--}$type bismuth telluride NWs with thicknesses of 7 nm (solid lines) and 15 nm (dashed lines) at the excess hole concentration ${p_{ex}=5 \times 10^{18} \: \text{cm}^{-3}}$. For the intrinsic (${\boldsymbol{p}\--}$type) NWs, the energy spectrum is calculated at gate voltages of 15 and 42 V (60 and 45 V) for the NW thicknesses 7 and 15 nm, respectively.}
\end{figure*}
In this section, we consider the temperature dependence of the thermoelectric parameters of the 7- and 15-nm-thick bismuth telluride NWs at the gate voltages corresponding to a maximum value of the figure of merit at room temperature. 

Figure~\ref{fig:7} presents the temperature dependence of the first three electron and hole energy subbands calculated by means of the SEM for the NWs with thicknesses of 7 and 15 nm. The subband energies seem to shift with temperature because the Fermi level is chosen as an origin in the band diagram. Actually, the NW Fermi level does shift with temperature. The shift of the Fermi level is mainly due to the change in the ratio between the numbers of carriers with opposite polarity with temperature.  At a low temperature, electrons dominate in the intrinsic NWs. For the 15-nm-thick (7-nm-thick) NW, the Fermi level overlaps the electron subbands, when the temperature is less than 190 K (110 K).  A number of holes increases with temperature; hence, the Fermi level tends to the middle of the NW band gap. For the 7-nm-thick (15-nm-thick) ${\boldsymbol{p}\--}$type NWs at the gate voltage of 60 V (45 V), the electrons (holes) dominate at a low temperature; hence, the Fermi level is close to the conduction (valence) subband. For the 15-nm-thick ${\boldsymbol{p}\--}$type NW at the excess hole concentration ${p_{ex}=5 \times 10^{18} \: \text{cm}^{-3}}$, the Fermi level overlaps the hole subbands when the temperature is less than 270 K. The separation between the energy subbands decreases with the NW thickness. For the 7-nm-thick (15-nm-thick) intrinsic NW, the band gap decreases from a value of 338 meV (189 meV) at a temperature of 77 K to a value of 321 meV (180 meV) at a temperature of 500 K. For the 7-nm-thick (15-nm-thick) ${\boldsymbol{p}\--}$type NW, the band gap of 493 meV (229 meV) does not depend on temperature. The splitting between the first and second subbands is much greater than the splitting between the second and third subbands. For the 7-nm-thick intrinsic NW, the splitting between the second and the third electron subbands increases from a value of 14 to 39 meV with temperature, while the corresponding hole subbands remain very close to each other. For the 7-nm-thick (15-nm-thick) ${\boldsymbol{p}\--}$type NW at the gate voltage of 60 V (45 V), the splitting between the second and third electron subbands of 57 meV (18 meV) is about 5 (3) times greater than the splitting between the corresponding hole subbands of 12 meV (6 meV). This effect is due to the difference between the electron and hole masses. We recall that,  for the nanowires, the electron density of states have peculiarities at the edge of energy subbands, which leads to the increase in both the carrier concentration and the Seebeck coefficient as well as the decrease in the thermal conductivity. Hence, for the ${\boldsymbol{n}\--}$type bismuth telluride NWs, the maximal value of the figure of merit is achieved at less value of the gate voltage in comparison with that for the ${\boldsymbol{p}\--}$type NWs because of the large separation between the second and third electron subbands.
\begin{figure}[h]
\includegraphics{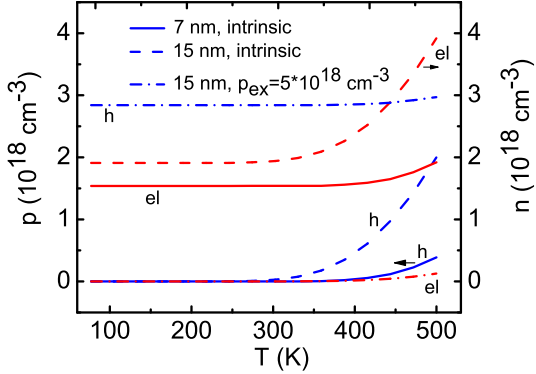}
\caption{\label{fig:8}(Color online) Temperature dependence of the electron and hole concentrations (a) for the intrinsic bismuth telluride NWs with thicknesses of 7 (solid line), 15 nm (dashed line) at gate voltages of 15 V and 42 V, correspondingly, and for the ${\boldsymbol{p}\--}$type bismuth telluride NWs with a thickness of 15 nm (dashed-dot line) at a gate voltage of 45 V, when ${p_{ex}=5 \times 10^{18} \: \text{cm}^{-3}}$.}
\end{figure}

Figure~\ref{fig:8} shows that, for the intrinsic NWs with thicknesses of 7 and 5 nm as well as for the ${\boldsymbol{p}\--}$type NW with a thickness of 15 nm at the excess hole concentration ${p_{ex}=5 \times 10^{18} \: \text{cm}^{-3}}$, the NW charge carrier concentration increases with temperature. While, for the ${\boldsymbol{p}\--}$type (${\boldsymbol{n}\--}$type) NW with a thickness of 7 nm at the excess hole (electron) concentration ${5 \times 10^{18} \: \text{cm}^{-3}}$ and ${1 \times 10^{19} \: \text{cm}^{-3}}$, the NW hole (electron) concentration  does not depend on temperature. It is equal to ${1.3 \times 10^{18} \: \text{cm}^{-3}}$ (${1.4 \times 10^{18} \: \text{cm}^{-3}}$) and ${2.3 \times 10^{18} \: \text{cm}^{-3}}$, respectively.  The temperature dependence of the carrier concentration decreases with both an increase in the excess hole (electron) concentration and a decrease of the NW thickness. Generally, the carrier concentration of the considered NWs does not depend on temperature till the room temperature. 
\begin{figure*}
\includegraphics{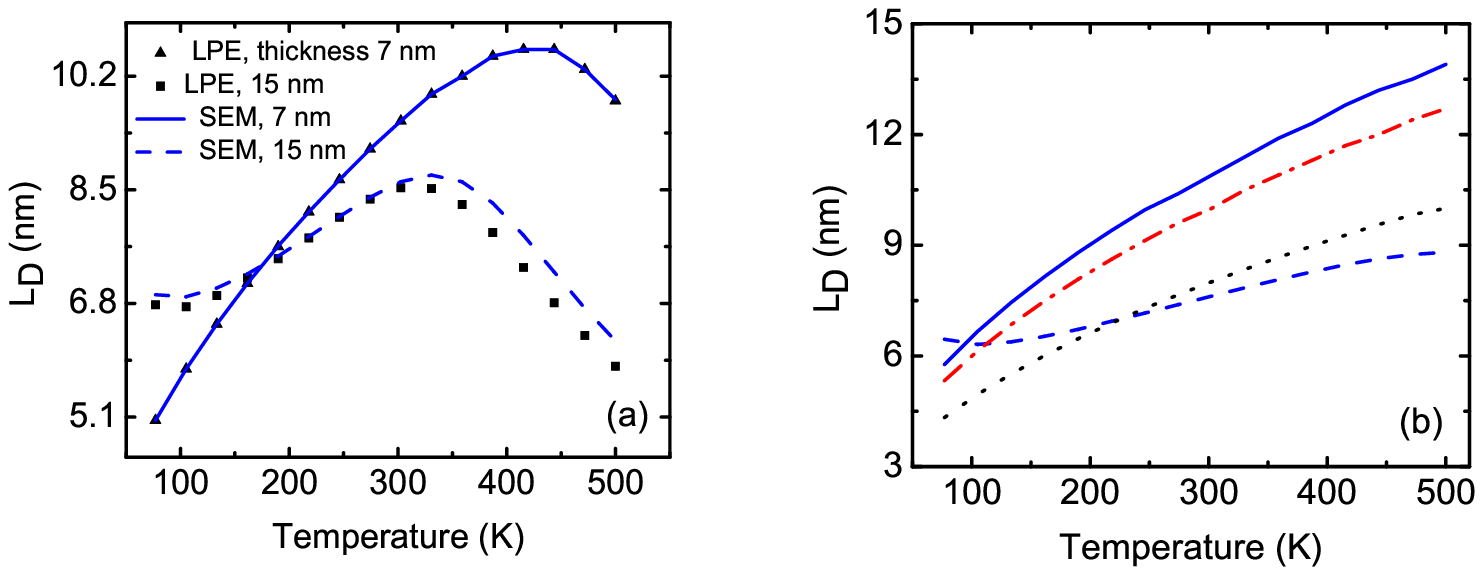}
\caption{\label{fig:9}(Color online) Temperature dependence of the Debye screening length (a) for the intrinsic bismuth telluride NWs with thicknesses of 7 (solid line) and 15 nm (dashed line) at gate voltages of 15 V and 42 V, correspondingly, and (b) for the ${\boldsymbol{p}\--}$type bismuth telluride NWs with thicknesses of 7 nm (solid line) and 15 nm (dashed line) at gate voltages of 60 and 45 V, respectively, when ${p_{ex}=5 \times 10^{18} \: \text{cm}^{-3}}$. At ${V_g =-35}$ V (${V_g =75}$ V), the dashed-dotted line (dotted line) corresponds to the Debye screening length for the ${\boldsymbol{n}\--}$type (${\boldsymbol{p}\--}$type) bismuth telluride NW with a thickness of 7 nm when ${n_{ex}=5 \times 10^{18} \: \text{cm}^{-3}}$ (${p_{ex}=1 \times 10^{19} \: \text{cm}^{-3}}$).}
\end{figure*}

\begin{figure*}
\includegraphics{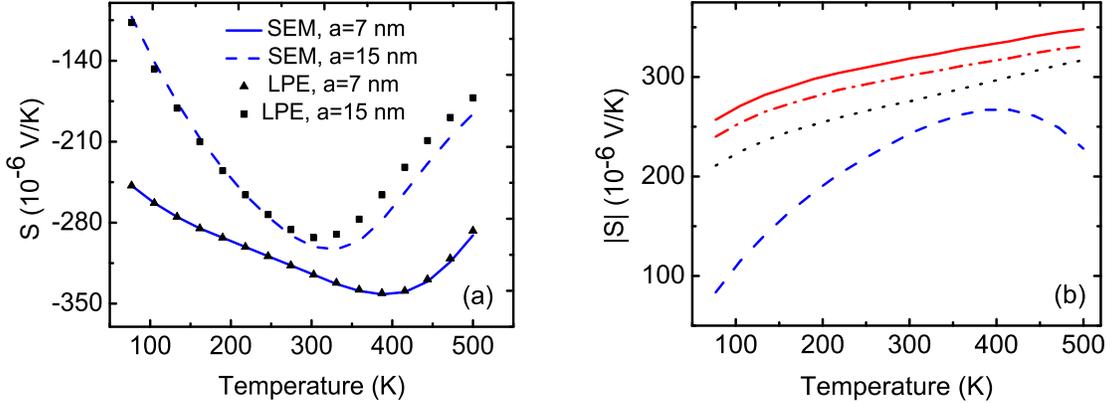}
\caption{\label{fig:10} (Color online) Temperature dependence (a) of the Seebeck coefficient ($S$) for the intrinsic bismuth telluride NWs with thicknesses of 7 nm (solid line) and 15 nm (dashed line) at gate voltages of 15 and 42 V, correspondingly, and (b) of the absolute values of $S$ for the ${\boldsymbol{p}\--}$type bismuth telluride NWs with thicknesses of 7 nm (solid line) and 15 nm (dashed line) at gate voltages of 60 and 45 V, respectively, when ${p_{ex}=5 \times 10^{18} \: \text{cm}^{-3}}$. At Vg =-35 V (Vg =75 V), the dashed-dotted line (dotted line) corresponds to the $S$ for the ${\boldsymbol{n}\--}$type (${\boldsymbol{p}\--}$type) bismuth telluride NW with a thickness of 7 nm when ${n_{ex}=5 \times 10^{18} \: \text{cm}^{-3}}$ (${p_{ex}=1 \times 10^{19} \: \text{cm}^{-3}}$).}
\end{figure*}

\begin{figure*}
\includegraphics{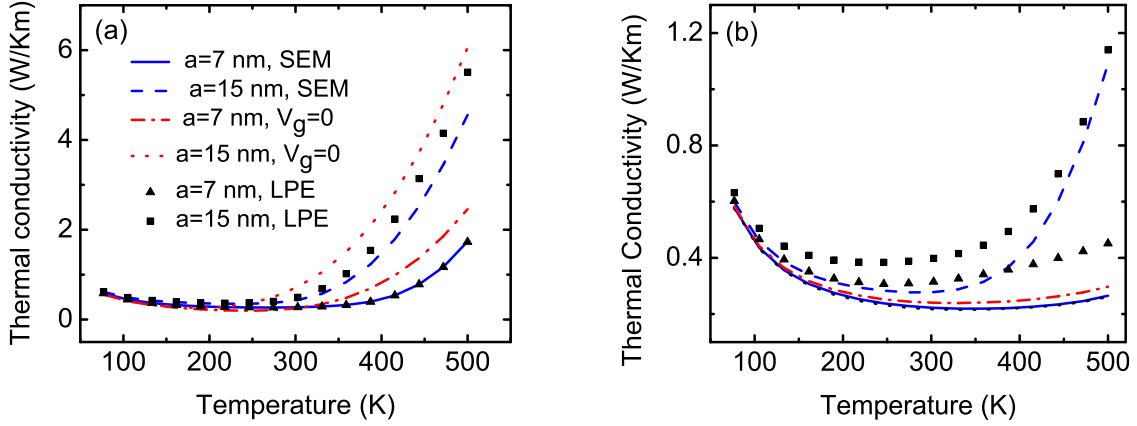}
\caption{\label{fig:11} (Color online) Temperature dependence of the thermal conductivity (a) for the intrinsic bismuth telluride NWs with thicknesses of 7 nm (solid line) and 15 nm (dashed line) at gate voltages of 15 and 42 V, correspondingly, and (b) for the ${\boldsymbol{p}\--}$type bismuth telluride NWs with thicknesses of 7 nm (solid line) and 15 nm (dashed line) at the gate voltages of 60 and 45 V, respectively, when ${p_{ex}=5 \times 10^{18} \: \text{cm}^{-3}}$. At ${V_g =-35}$ V (${V_g =75}$ V), the dashed-dotted line (dotted line) corresponds to the thermal conductivity for the ${\boldsymbol{n}\--}$type (${\boldsymbol{p}\--}$type) bismuth telluride NW with a thickness of 7 nm when ${n_{ex}=5 \times 10^{18} \: \text{cm}^{-3}}$ (${p_{ex}=1 \times 10^{19} \: \text{cm}^{-3}}$). At the zero gate voltage, the dependence of the thermal conductivity is shown by the dashed-dotted (filled triangles) and dotted lines (filled squares) for the intrinsic (${\boldsymbol{p}\--}$type) NWs with thicknesses of 7 and 15 nm, respectively, in Fig.~\ref{fig:11}(a) (Fig.~\ref{fig:11}(b)). The filled squares (triangles) correspond to the data obtained in the LPE approach for the NW with a thickness of 15 nm (7 nm) (see Fig.~\ref{fig:11}(a)).}
\end{figure*}

\begin{figure*}
\includegraphics{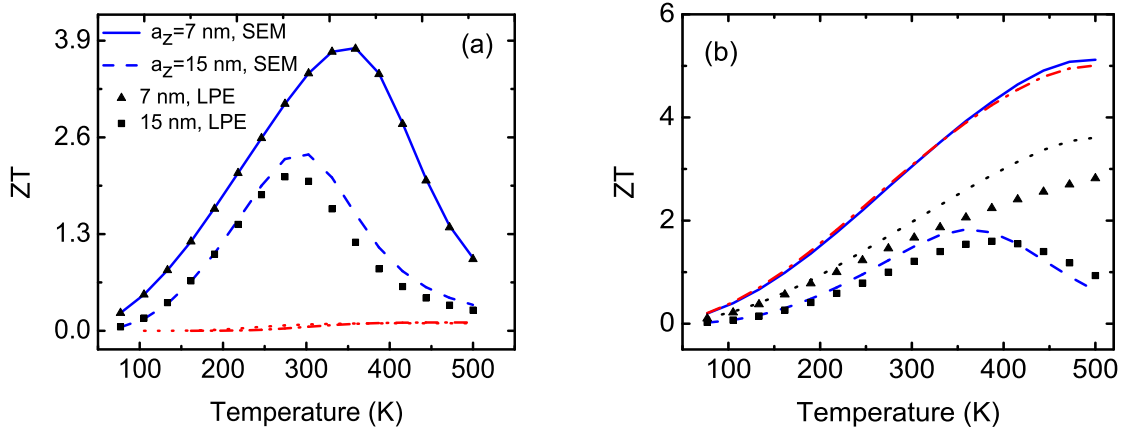}
\caption{\label{fig:12} (Color online) Temperature dependence of the figure of merit (a) for the intrinsic bismuth telluride NWs with thicknesses of 7 nm (solid line) and 15 nm (dashed line) at gate voltages of 15 and 42 V, correspondingly, and (b) for the ${\boldsymbol{p}\--}$type bismuth telluride NWs with thicknesses of 7 nm (solid line) and 15 nm (dashed line) at the gate voltages of 60 V and 45 V, respectively, when ${p_{ex}=5 \times 10^{18} \: \text{cm}^{-3}}$. At ${V_g =-35}$ V (${V_g =75}$ V), the dashed-dotted line (dotted line) corresponds to the figure of merit for the ${\boldsymbol{n}\--}$type (${\boldsymbol{p}\--}$type) bismuth telluride NW with a thickness of 7 nm when ${n_{ex}=5 \times 10^{18} \: \text{cm}^{-3}}$ (${p_{ex}=1 \times 10^{19} \: \text{cm}^{-3}}$). At the zero gate voltage, the dependence of the figure of merit is shown by the dashed-dotted (filled triangles) and dotted lines (filled squares) for the intrinsic (${\boldsymbol{p}\--}$type) NWs with thicknesses of 7 and 15 nm, respectively, in Fig.~\ref{fig:12}(a) (Fig.~\ref{fig:12}(b)). The filled squares (triangles) correspond to the data obtained in the LPE approach for the NW with a thickness of 15 nm (7 nm) (see Fig.~\ref{fig:12}(a)).}
\end{figure*}

Both the SEM and LPE approach are used to obtain the data presented in Figs~\ref{fig:9}-\ref{fig:12}. Figure~\ref{fig:9} shows the temperature dependence of the Debye screening length (${L_D}$) for the 7- and 15-nm-thick  NWs. Since the electron (hole) gas in the 15-nm-thick intrinsic (${\boldsymbol{p}\--}$type) NW is degenerate when the temperature is less than 150 K (see Fig.~\ref{fig:7}(a)), the Debye screening length does not change with temperature. For the doped NWs, the Debye screening length monotonically increases with temperature when the Fermi level approaches the middle of the band gap. For the 7-nm-thick NWs, the ${L_D}$ increases by 2.3 times, while it increases by 40 per cent for the 15-nm-thick ${\boldsymbol{p}\--}$type NW. In contrast, the temperature dependence of the ${L_D}$ for the intrinsic NWs is not monotonic.  At a high temperature, the electron concentration increases with temperature. Hence, the Debye screening length decreases. For the 7- and 15-nm-thick intrinsic NWs, the ${L_D}$ increases by 2 times and by 25 per cent, respectively. The Debye screening length also decreases with an increase in the excess hole concentration. For example, for the 7-nm-thick ${\boldsymbol{p}\--}$type NW with the excess hole concentration ${p_{ex}=1 \times 10^{19} \: \text{cm}^{-3}}$, the ${L_D}$ is less than that for the 7-nm-thick NW at ${p_{ex}=5 \times 10^{18} \: \text{cm}^{-3}}$ by 25 per cent in the whole temperature range. Therefore, the Debye screening length strongly depends on temperature for the NWs with a small thickness.

Figure~\ref{fig:10} shows the temperature dependences of the Seebeck coefficient for the NWs with thicknesses of 7 and 15 nm. For the intrinsic NWs and the 15-nm-thick ${\boldsymbol{p}\--}$type NW, the temperature dependence of the Seebeck coefficient ($S$) is not monotonic, because the electron (hole) concentration increases at a high temperature. For both the ${\boldsymbol{n}\--}$ and ${\boldsymbol{p}\--}$type NWs with a thickness of 7 nm, at gate voltages of -35 and 60 V, respectively, the Seebeck coefficient is negative, because the electrons dominate. While, for the 7- and 15-nm-thick ${\boldsymbol{p}\--}$type NWs with the excess hole concentrations of ${p_{ex}=1 \times 10^{19} \: \text{cm}^{-3}}$ and  ${p_{ex}=5 \times 10^{18} \: \text{cm}^{-3}}$, correspondingly, at  gate voltages of 75 and 45 V, the $S$ is positive owing to the domination of holes in the system. For the 7-nm-thick (15-nm-thick) intrinsic and 15-nm-thick ${\boldsymbol{p}\--}$type NWs, the maximum absolute values of the Seebeck coefficient 342 (302) and 267  V/K are achieved at a temperature of 400 (325) and 390 K, correspondingly. At the room temperature, for the 7-nm-thick (15-nm-thick) intrinsic NWs, the absolute value of the Seebeck coefficient 325  V/K (301  V/K) at the gate voltage of 15 V (42 V) is greater than a value of 252   V/K (167  V/K) at the zero gate voltage by 29 (80) per cent. For the 7-nm-thick ${\boldsymbol{p}\--}$type NW at the excess hole concentration ${p_{ex}=5 \times 10^{18} \: \text{cm}^{-3}}$ (${p_{ex}=1 \times 10^{19} \: \text{cm}^{-3}}$), the absolute value of $S$ monotonically increases with temperature by 35 (50) per cent.  For the 7-nm-thick ${\boldsymbol{n}\--}$type NW, ${|S|}$ is less than that for the 7-nm-thick ${\boldsymbol{p}\--}$type NW by 5 per cent in the whole temperature interval. The monotonic behavior of the Seebeck coefficient as a function of temperature results from the weak local variation of the electron concentration, when the Debye screening length is greater than the NW thickness (see Fig.~\ref{fig:9}(b)). At the room temperature, for the 7-nm-thick ${\boldsymbol{p}\--}$type (${\boldsymbol{n}\--}$type) NWs with the excess hole (electron) concentration ${p_{ex}=5 \times 10^{18} \: \text{cm}^{-3}}$ (${n_{ex}=5 \times 10^{18} \: \text{cm}^{-3}}$) and ${p_{ex}=1 \times 10^{19} \: \text{cm}^{-3}}$, the absolute value of the Seebeck coefficient 319 (302) and 276  V/K at gate voltage of 60 (-35)  and 75 V, respectively, is greater than a value of 206 (187) and 142  V/K at the zero gate voltage by 54 (61) and 95 per cent, correspondingly. For the 15-nm-thick ${\boldsymbol{p}\--}$type NW at the room temperature, the Seebeck coefficient of 196  V/K increases by 25 per cent when applying the gate voltage of 45 V. Hence, the dependence of the Seebeck coefficient on temperature is diminished with a decrease in the NW thickness, while the absolute value of $S$ increases under the applied electric field.

Figure~\ref{fig:11} presents the temperature dependence of the thermal conductivity of the bismuth telluride NWs with thicknesses of 7 and 15 nm. For comparison, we depict the temperature dependence of the thermal conductivity for the NWs with the same thicknesses at a zero gate voltage. For the 7- and 15-nm-thick intrinsic NWs, the thermal conductivity at the gate voltages of 15 and 42 V, respectively, is less than that at the zero gate voltage when the temperature is greater than 310 and 240 K. At the room temperature, for the 15-nm-thick (7-nm-thick) intrinsic NW, the thermal conductivity of 0.434 W/Km (0.274 W/Km) at the gate voltage of 40 V (15 V) is less (greater) compared to the value of 0.727 W/Km (0.254 W/Km) at the zero gate voltage by 40 (8) per cent. The temperature dependence of the thermal conductivity is not monotonic because its electron (hole) part increases with temperature, while its lattice part decreases. At a high temperature, for the intrinsic NWs, the temperature dependence of the thermal conductivity is much stronger in comparison with that for the doped NWs. The increase in the thermal conductivity of the intrinsic NWs arises from the additional thermal transport channel produced by the electron-hole pairs (see Eq.~\ref{eq:10}). The electric field bias enables the charge carrier concentration to increase. As a result, for the 7-nm-thick (15-nm-thick) intrinsic NW at the gate voltage of 15 V (42 V), the minimum value of the thermal conductivity of 0.268 W/Km (0.352 W/Km) at ${T=275}$ K (247 K) is greater than a value of 0.195 W/Km (0.259 W/Km) at the zero gate voltage by 36 per cent, when ${T=247}$ K (190 K).  For the doped NWs, the applied lateral electric field decreases the thermal conductivity in the whole temperature range. For the 7-nm-thick (15-nm-thick) ${\boldsymbol{p}\--}$type NW at the gate voltage of 60 V (45 V), the minimum value of the thermal conductivity of 0.219 W/Km (0.277 W/Km) at ${T=360}$ K (275 K) is less than the value of 0.307 W/Km (0.384 W/Km) at the zero gate voltage by 27 per cent when ${T=246}$ K.  At the room temperature, for the 7-nm-thick (15-nm-thick) ${\boldsymbol{p}\--}$type NW at the excess hole concentration ${p_{ex}=5 \times 10^{18} \: \text{cm}^{-3}}$, the thermal conductivity of 0.222 W/Km (0.278 W/Km) at the gate voltage of 60 V (45 V) is less than a value of 0.315 W/Km (0.398 W/Km) at the zero gate voltage by 30 per cent. For the 7-nm-thick ${\boldsymbol{p}\--}$type NW, the temperature dependence of the thermal conductivity is identical to that for the ${\boldsymbol{n}\--}$type NW.  Therefore, the applied electric field permits decreasing the NW thermal conductivity. 

Figure~\ref{fig:12} shows the temperature dependence of the figure of the merit for the intrinsic and doped NWs with thicknesses of 7 and 15 nm. At the gate voltage of 15 V (42 V), the intrinsic 7-nm-thick (15-nm-thick) NW exhibits the maximum value of the figure of merit as high as 3.8 (2.4) at the temperature of ${T=360}$ K (300 K). As a result of the great increase in the thermal conductivity and decrease in the Seebeck coefficient for the intrinsic 7-nm-thick (15-nm-thick) NW within the high temperature range, the figure of merit decreases rapidly with temperature when ${T>360}$ K (300 K). For the 7-nm-thick (15-nm-thick) ${\boldsymbol{p}\--}$type NWs with the excess hole concentration ${p_{ex}=5 \times 10^{18} \: \text{cm}^{-3}}$, the maximum value of the figure of merit is 5.12 (1.83) at a temperature of 500 K (360 K), when the gate voltage is 60 V (45 V). As a function of temperature, the figure of merit for the 7-nm-thick ${\boldsymbol{n}\--}$type NW resembles that for the ${\boldsymbol{p}\--}$type NW. Their figure of merit is about twice as large as that at the zero gate voltage. For the 7-nm-thick doped NWs, the monotonic dependence of the figure of merit on temperature is attributed to the slight temperature dependence of the Seebeck coefficient and the thermal conductivity. As a result of decrease of the Seebeck coefficient with an increase in the excess hole concentration, the figure of merit decreases. For example, at the temperature of 500 K, for the 7-nm-thick ${\boldsymbol{p}\--}$type NW with the excess hole concentration ${p_{ex}=1 \times 10^{19} \: \text{cm}^{-3}}$, the figure of merit of 3.61 is less than that corresponding to the NW with the excess hole concentration ${n_{ex}=5 \times 10^{18} \: \text{cm}^{-3}}$ by 30 per cent. For the intrinsic NWs, the extreme point of the figure of merit is close to the room temperature. Both the confinement effect and the excess holes (electrons) produce a shift of the extreme point toward a high temperature. Hence, for the intrinsic NWs, the electric field effect gives rise to an increase in the figure of merit by an order of magnitude; while, for the doped NWs, the figure of merit can increase as much as twice.
\section{\label{sec:level6} CONCLUSIONS}
In this paper, we showed that a combination of the quantum confinement effect in the bismuth telluride quantum wires with a properly applied electric bias can increase the thermoelectric power conversion by an order of magnitude. For the considered bismuth telluride NWs, the electron-electron exchange-correlation interaction is negligible because of the small electron (hole) concentration. At room temperature, the exchange-correlation energy does not exceed the electron thermal energy. The EFE can effectively control the type of the NW conductivity by adjusting the charge carrier concentration.  It compares favorably with the doping effect because the number of scattering centers is not changed under an the electric field bias. The variation in the charge carrier concentration leads to a self-consistent change of the size quantization effect. For bismuth telluride NWs, the electron effective mass is less than the hole effective mass. As a result, the splitting between the electron subbands is greater than that between the hole subbands. In addition, for bismuth telluride material, the electron mobility is about twice as high as the hole mobility. It indicates that the thermoelectric properties of the ${\boldsymbol{n}\--}$type NWs are superior to those of the ${\boldsymbol{p}\--}$type NWs. The dependence of the Seebeck coefficient on temperature is diminished with a decrease in the NW thickness, while the absolute value of $S$ increases under the applied electric filed. In dependence on the NW thickness and excess hole (electron) concentration, the electric field bias enables the Seebeck coefficient to increase by 30-90 per cent. For the 7-nm-thick (15-nm-thick) intrinsic and 15-nm-thick ${\boldsymbol{p}\--}$type NWs, under the applied electric field, the maximum absolute values of the Seebeck coefficient of 342 (302) and 267  V/K are achieved at a temperature of 400 (325) and 390 K, correspondingly. The confinement effect on the electron density of states leads to an increase in the NW Seebeck coefficient, while the increase in the carrier concentration causes the decrease in $S$. The applied electric field leads to a decrease in the electronic contribution to the NW thermal conductivity. For example, at the room temperature, for the 15-nm-thick ${\boldsymbol{p}\--}$type (intrinsic) NW with the excess hole concentration ${p_{ex}=5 \times 10^{18} \: \text{cm}^{-3}}$, the thermal conductivity of 0.278 W/Km (0.426 W/Km) at gate voltage of 45 V (42 V)  is less  compared to the respective value of 0.398 W/Km (0.727 W/Km) at the zero gate voltage by 30 (40) per cent. For the 7-nm-thick (15-nm-thick) ${\boldsymbol{p}\--}$type NWs with the excess hole concentration ${p_{ex}=5 \times 10^{18} \: \text{cm}^{-3}}$, the maximum value of the figure of merit is 5.12 (1.83) at a temperature of 500 K (360 K), when the gate voltage is 60 V (45 V). At the room temperature, for the 7-nm-thick (15-nm-thick) ${\boldsymbol{p}\--}$type NW with the excess hole concentration of ${p_{ex}=5 \times 10^{18} \: \text{cm}^{-3}}$, the maximum value of the figure of merit of 3.1 (1.5) is greater than the corresponding value ${ZT=1.65}$ (1.19) at the zero gate voltage by 90 (25) per cent. For the 7-nm-thick (15-nm-thick) intrinsic NW, the maximum value of the figure of merit is as high as 3.4 (2.4) at the gate voltage of 15 V (42 V). It is an order of magnitude greater than the corresponding value of the figure of merit at the zero gate voltage.  
\appendix
\section{Linearization of Poisson's equation}
In the semi-classical approximation, to take into account the electron-screening effects, one can write the local variation of the electron and hole concentrations of each valley in a nanowire as \cite{Ridley, Mitin}
\begin{equation}
n_{1D}(x,z)=\frac{N^{1D}_{c,v}}{a_x a_z} \sum_{n=1}^{n_{max}} \sum_{l=1}^{l_{max}(n)} \Phi_{-1/2} \left[ \eta_{n, l} \pm \frac{e \varphi(x,z)}{k_B T} \right].
\label{eq:A1}
\end{equation}
Here, the factor ${N^{1D}_{c,v}=\left( 2m^{e,h}_{y} k_B T/ \pi \hbar^2 \right)^{1/2}}$  denotes the effective density of electron (hole) states. Sign "+" ("-") corresponds to electrons (holes). The reduced chemical potentials  ${\eta^{c(v)}_{n, l}}$ for an electron subband and a hole subband are defined as
\begin{equation}
\eta^{c}_{n, l}=\frac{E_F-E_c-E^{e}_{n, l}}{k_B T},
\label{eq:A2}
\end{equation}
\begin{equation}
\eta^{v}_{n, l}=\frac{E_v+E^{h}_{n, l}-E_F}{k_B T}.
\label{eq:A3}
\end{equation}
For a non-degenerate semiconductor material, if the applied electric filed is small enough compared to the difference between the Fermi energy and the bottom (top) of the conduction (valence) band such as  ${\left| E_F-E_{c,v}-E^{e,h}_{n,l} \right|>> \left| e\varphi(x,z) \right|}$, then the electron (hole) concentration ${n_{1D}}$ (${p_{1D}}$) can be expanded in Taylor series in terms of  ${e \varphi/k_B T}$. Using this expansion, the right hand side of Eq.~(\ref{eq:2}) can be linearized with regard to the electrostatic potential ${\varphi}$. Hence, Poisson's equation can be rewritten in the linearized form as
\begin{equation}
\Delta \varphi(x,z)=\frac{\varphi(x,z)}{L^{2}_{D}}.
\label{eq:A4}
\end{equation}
Here, the Debye screening length is defined by
\begin{widetext}
\begin{equation}
L_{D}=\left[ \frac{e^2}{a_x a_z \epsilon_{Bi_2Te_3} \epsilon_0 k_B T} \sum_{n=1}^{n_{max}}\sum_{l=1}^{l_{max}(n)}{ \left( N^{1D}_{v} \frac{\text{d}\Phi_{-1/2}(\eta^{v}_{n,l})}{\text{d}\eta^{v}_{n,l}} + N^{1D}_{c} \frac{\text{d}\Phi_{-1/2}(\eta^{c}_{n,l})}{\text{d}\eta^{c}_{n,l}} \right)} \right]^{-1/2}.
\label{eq:A5}
\end{equation}
\end{widetext}
Since the electric field is directed along the $z$ axis and the exchange-correlation energy ${V_{xc}}$ is negligible in case of the intrinsic NWs, the variables in Eqs.~(\ref{eq:1}) and ~(\ref{eq:A4}) can be separated. Therefore, the analytical solution of the boundary problem~(\ref{eq:6}),~(\ref{eq:7}), and ~(\ref{eq:A4}) reads
\begin{equation}
\varphi(z)=V_g \frac{L_D \epsilon_{SiO_2} \text{sinh}(z/L_D)}{d_{SiO_2} \epsilon_{Bi_2Te_3} \text{cosh}(a_z/L_D)}.
\label{eq:A6}
\end{equation}
The solution of the Schr\"{o}dinger equation for electrons reads 
\begin{equation}
\Psi_{n,l}(x,z)=\sqrt{\frac{2}{a_x}}\text{sin}(\pi l x/a_x)Z_n(z).
\label{eq:A7}
\end{equation}
The unknown function ${Z_n(z)}$ satisfies the following equation 
\begin{equation}
-\frac{\hbar^2}{2m_z}\frac{\text{d}^2 Z_n}{\text{d} z^2} - e \varphi(z)Z_n(z)=\lambda_n Z_n(z),
\label{eq:A8}
\end{equation}
and the corresponding boundary conditions ${Z_n(0)=Z_n(a_z)=0}$. The electron subband energy is given by the expression ${E^{e}_{n,l}=\lambda_n + \hbar^2 \pi^2 l^2/2 m^{e}_{x} a^{2}_{x}}$. So, we recast the self-consistent system of differential equations~(\ref{eq:1}) and (\ref{eq:2}) into the simplified self-consistent system of  one differential equation~(\ref{eq:A8}) and two mathematical expressions~(\ref{eq:A5}) and (\ref{eq:A6}) taking into account charge neutrality equation~(\ref{eq:4}). To solve the last system of equations, we apply the following algorithm: STEP 1: Initial guess of the Debye screening length ${L^{(0)}_{D}}$. STEP 2:  Calculation of the electrostatic potential ${\varphi^{(k)}(z)}$ by means of Eq.~(\ref{eq:A6}) using  ${L^{(k)}_{D}}$. STEP 3: Solving Eq.~(\ref{eq:A8}) for the eigenvalues  ${\lambda^{(k)}_{n}}$  using  ${\varphi^{(k)}(z)}$. STEP 4: Solving Eq.~(\ref{eq:4}) for the Fermi energy ${E^{(k)}_{F}}$ using ${E^{(k)}_{n,l}}$. STEP 5:  Calculation of ${L^{(k+1)}_{D}}$  by means of Eq.~(\ref{eq:A5}) using ${E^{(k)}_{n,l}}$. STEP 6: IF  ${\left| L^{(k+1)}_{D}-L^{(k)}_{D} \right| / L^{(k+1)}_{D}>\delta}$ THEN  ${L^{(k)}_{D}:= \left( L^{(k+1)}_{D}+L^{(k)}_{D} \right) /2 }$  AND GOTO STEP 2 ELSE STOP. The solution of Eq.~(\ref{eq:A8}) can be obtained analytically and numerically. To find the analytical solution, one transforms Eq.~(\ref{eq:A8})  into the modified Mathieu equation.\cite{McLachlan, Bejenari_SST19} Therefore, in case of intrinsic semiconductor NWs, the linearized Poisson equation (LPE) approach allows us to simplify solving the self-consistent system of Schr\"{o}dinger and Poisson equations significantly. To solve equation~(\ref{eq:A8}) numerically, we can use the spectral method or spectral element method. \cite{Cheng_JCE3, Boyd}
\section{Spectral  Method with trigonometric basis functions}
The function ${Z(z)}$ in Eq.~(\ref{eq:A8}) can be decomposed in terms of the trigonometric basis functions
\begin{equation}
Z(t)=\sum_{k=1}^{N}c_k \vartheta_k(t).
\label{eq:B1}
\end{equation}
Here, we set the reduced variable ${t=z/L_D}$ which ranges in the interval ${[0, r]}$. The upper limit of integration is ${r=a_z/L_D}$. The basis ${\left \{ \vartheta_k \right\}^{N}_{k=1} }$  consists of the wave functions ${\vartheta_k(t)=\sqrt{2/r}\:\text{sin}(\pi k t/r)}$  of an electron in the one-dimensional Infinite Square well potential. To find the unknown constants ${c_k}$ we substitute expression ~(\ref{eq:B1}) into Eq.~(\ref{eq:A8}), multiply both sides of the equation by the conjugated function  ${\vartheta^{*}_{m}(t)}$ and integrate from 0 to $r$. Finally, we obtain a system of linear algebraic equations for $N$ unknowns ${c_k}$. The eigenvalues ${\lambda_n}$ can be calculated by means of the secular (characteristic) equation 
\begin{widetext}
\begin{equation}
\left| \left[ \left( \frac{\pi k}{r} \right)^2 - \tilde{\lambda} \right ] \delta_{m,k}- \sum_{k=1}^{N}\frac{2m_z L^{2}_{D}}{\hbar^2} e \int_{0}^{r} \vartheta^{*}_{m}(t) \varphi(t) \vartheta_k (t)dt \right |=0,
\label{eq:B2}
\end{equation}
\end{widetext}
where $m$ and ${n=1, 2, \ldots N}$ and ${\tilde{\lambda}=2 \lambda m_z L^{2}_{D} / \hbar^2}$. We take ${N=30}$ here. The integral in Eq.~(\ref{eq:B2}) is calculated analytically when the electrostatic potential  ${\varphi(t)}$  is given by expression~(\ref{eq:A6}). In the general case, if the potential in Schr\"{o}dinger equation~(\ref{eq:1}) presents a complex function, we can solve this equation by using the more sophisticated method, that is, spectral element method described in the next subsection. 
\section{Spectral element method based on the Gauss-Lobatto-Legendre polynomials}
In the SEM approximation, the component of the electron wave function   and the electrostatic potential   are expanded in terms of the field variables \cite{Cheng_JCE3} 
\begin{equation}
Z(\tilde{z})=\sum_{j=0}^{N}Z(\tilde{z}_j) b_j(\tilde{z}),
\label{eq:C1}
\end{equation}
\begin{equation}
\varphi(\tilde{z})=\sum_{j=0}^{N} \varphi(\tilde{z}_j) b_j(\tilde{z}),
\label{eq:C2}
\end{equation}
where the reduced variable ${\tilde{z}}$  ranges in the interval ${-1 \leq \tilde{z} \leq 1}$. The field variables in the Schr\"{o}dinger and Poisson's equations are represented by the Gauss-Lobatto-Lelendre polynomials ${b_j(\tilde{z})}$ of the $N$th order
\begin{equation}
b_j(\tilde{z})=\frac{(1-\tilde{z}^2)L'_{N}(\tilde{z})}{N(N+1)L_N(\tilde{z}_j) (\tilde{z}_j-\tilde{z})},
\label{eq:C3}
\end{equation}
where ${L_N(z)}$ is the $N$th order Legendre polynomial. The $j$th zero ${\tilde{z}_j}$  is a root of the equation  ${L'_N(\tilde{z}_j)=0}$. We suppose $N$=15 here. To obtain a weak form of the Schr\"{o}dinger and Poisson's equations, we substitute expressions~(\ref{eq:C1}) and~(\ref{eq:C2}) into the modified Eqs.~(\ref{eq:1}) and ~(\ref{eq:2}), multiply both sides of these equations by the function ${b_j(\tilde{z})}$  and integrate from -1 to 1. Furthermore, we apply the Gaussin-Legendre quadrature ${w_l=2/(1-\tilde{z}^{2}_{l})L^{2}_{N+1}(\tilde{z}_l)}$  to evaluate the integration and finally obtain the discrete Schr\"{o}dinger and Poisson's equations as 
\begin{widetext}
\begin{eqnarray}
\sum_{j=0}^{N}\sum_{l=0}^{N}w_l \left\{ \frac{1}{J} \frac{\text{d} b_i(\tilde{z}_l)}{\text{d} \tilde{z}} \frac{\text{d} b_j(\tilde{z}_l)}{\text{d} \tilde{z}} + \frac{2J m_z a^{2}_{z}}{\hbar^2} \left[  V_{xc}(\tilde{z}_l) + E_c - e \varphi(\tilde{z}_l) \right ] b_i(\tilde{z}_l) b_j(\tilde{z}_l) \right \} Z(\tilde{z}_j)  \nonumber \\
 = \sum_{j=0}^{N}\sum_{l=0}^{N} w_l \tilde{\lambda} J b_i(\tilde{z}_l) b_j(\tilde{z}_l)  Z(\tilde{z}_j),  
\label{eq:C4}
\end{eqnarray}
\begin{equation}
\sum_{j=0}^{N}\sum_{l=0}^{N} \epsilon_0 \epsilon_{Bi_2Te_3} \frac{w_l}{J} \frac{\text{d} b_i(\tilde{z}_l)}{\text{d} \tilde{z}} \frac{\text{d} b_j(\tilde{z}_l)}{\text{d} \tilde{z}} \varphi(\tilde{z}_j) = \sum_{l=0}^{N} a^{2}_{z} w_l J b_i(\tilde{z}_l) \rho(\tilde{z}_l),
\label{eq:C5}
\end{equation}
\end{widetext}
where ${i=0, 1,\ldots, N}$ , ${\tilde{\lambda}=2 \lambda m_z a^{2}_{z} / \hbar^2}$, and $J$ is the Jacobian. Eq.~(\ref{eq:C4}) can be solved by a generalized eigenvalue solver. However, because the mass matrix in Eq.~(\ref{eq:C4}) is diagonal, this equation can be converted into a more efficient regular eigenvalue problem.\cite{Cheng_JCE3} Eq.~(\ref{eq:C5}) can be solved by the Newton-Raphson method using the predictor-corrector approach.\cite{Press, Trellakis_JAP81, Paceli_IEEE} To solve the self-consistent system of Schrodiger and Poison's equations~(\ref{eq:C4}) and~(\ref{eq:C5}), we use the following algorithm: STEP 1: Suppose the electron-electron exchange-correlation energy is  ${V^{(0)}_{xc}(\tilde{z})=0}$. STEP 2:  Solving Poisson's equation~(\ref{eq:2}) with boundary conditions~(\ref{eq:6}) and~(\ref{eq:7}) for the electrostatic potential  ${\varphi^{(0)}(\tilde{z})}$, when ${\rho = 0}$. STEP 3: Solving Schr\"{o}dinger equation~(\ref{eq:C4}) for the eigenvectors ${Z^{(k)}_{n}(\tilde{z}_j)}$  and eigenvalues ${\tilde{\lambda}}^{(k)}_{n}$  taking into account the electrostatic potential ${\varphi^{(k)}(\tilde{z}_j)}$  and the electron-electron exchange-correlation energy  ${V^{(k)}_{xc}(\tilde{z}_j)}$. STEP 4: Solving Eq.~(\ref{eq:4}) for the Fermi energy ${E^{(k)}_{F}}$  using ${E^{(k)}_{n,l}}$. STEP 5: Solving Eq.~(\ref{eq:C5}) for ${\varphi^{(k)}(\tilde{z}_j)}$  by means of the Newton-Raphson metod using the predictor-corrector approach. STEP 6: Calculation of the electron-electron exchange-correlation energy ${V^{(k)}_{xc}(\tilde{z}_j)}$ by means of Eq.~(\ref{eq:3}) using the Fermi energy  ${E^{(k)}_{F}}$, energy spectrum  ${E^{(k)}_{n,l}}$, and wave functions ${\Psi^{(k)}_{n,l}(\tilde{z}_j)}$. STEP 7: IF ${\sum_{j=0}^{N} \left[ |\varphi^{(k+1)}(\tilde{z}_j)|- |\varphi^{(k)}(\tilde{z}_j)| \right ] ^2 / \sum_{j=0}^{N}|\varphi^{(k)}(\tilde{z}_j)|^2 > \delta}$  THEN  ${\varphi^{(k+1)}(\tilde{z}_j):=\varphi^{(k)}(\tilde{z}_j)}$  AND GOTO STEP 3 ELSE STOP.

\begin{acknowledgments}
I.B. acknowledges financial support of the Fulbright Scholar Program. I.B. also thanks Prof. A.A. Balandin for his kind invitation to join his research group at the Department of Electrical Engineering, UCR from September 1, 2008 to May 31, 2009. 
\end{acknowledgments}


\end{document}